\documentclass[fleqn,3p,onecolumn,11pt,nofootinbib,showkeys,showpacs]{revtex4-1}

\usepackage[small,justification=centerlast]{caption}
\usepackage{graphicx}
\usepackage{bm}
\usepackage{mathrsfs}
\usepackage[latin1]{inputenc}
\usepackage{array}
\usepackage{psfrag}
\usepackage{dsfont}
\usepackage{epsfig}
\usepackage{titletoc}
\usepackage{float}   
\usepackage{wrapfig} 
\usepackage{amsmath}\usepackage{amssymb,stmaryrd}
\usepackage{array}
\usepackage{epsfig}
\usepackage{cancel}
\usepackage[dvips]{feynmp}
\usepackage{upgreek}
\usepackage{subfig}
\usepackage{tabularx}
\usepackage{xcolor}
\usepackage{units}
\usepackage{textcomp}
\usepackage{wasysym}
\usepackage{amsthm}

\usepackage{textfit} 

\usepackage{hyperref}

\renewcommand{\eqref}[1]{\mbox{Eq.~(\ref{#1})}}
\newcommand{\tabref}[1]{\mbox{Tab.~\ref{#1}}}

\newcommand{\secref}[1]{\mbox{Sec.~\ref{#1}}}

\newcommand{\appref}[1]{\mbox{App.~\ref{#1}}}

\begin{document}

\title{First-order classical Lagrangians \\ for the nonminimal Standard-Model Extension}

\author{J.A.A.S. Reis} \email{jalfieres@gmail.com}
\affiliation{Departamento de F\'{i}sica, Universidade Federal do Maranh\~{a}o \\
65080-805, S\~{a}o Lu\'{i}s, Maranh\~{a}o, Brazil}
\author{M. Schreck} \email{marco.schreck@ufma.br}
\affiliation{Departamento de F\'{i}sica, Universidade Federal do Maranh\~{a}o \\
65080-805, S\~{a}o Lu\'{i}s, Maranh\~{a}o, Brazil}

\begin{abstract}
In this paper, we derive the general leading-order classical Lagrangian covering all fermion operators of the nonminimal Standard-Model Extension (SME). Such a Lagrangian is considered to be the point-particle analog of the effective field theory description of Lorentz violation that is provided by the SME. First of all, a suitable \textit{Ansatz} is made for the Lagrangian of the spin-degenerate operators $\hat{a}$, $\hat{c}$, $\hat{e}$, and $\hat{f}$ at leading order in Lorentz violation. The latter is shown to satisfy the set of five nonlinear equations that govern the map from the field theory to the classical description. After doing so, the second step is to propose results for the spin-nondegenerate operators $\hat{b}$, $\hat{d}$, $\hat{H}$, and $\hat{g}$. Although these are more involved than the Lagrangians for the spin-degenerate ones, an analytical proof of their validity is viable, nevertheless. The final step is to combine both findings to produce a generic Lagrangian for the complete set of Lorentz-violating operators that is consistent with the known minimal and nonminimal Lagrangians found in the literature so far. The outcome reveals the leading-order structure of the classical SME analog. It can be of use for both phenomenological studies of classical bodies in gravitational fields and conceptual work on explicit Lorentz violation in gravity. Furthermore, there may be a possible connection to Finsler geometry.
\end{abstract}

\maketitle

\newpage
\setcounter{equation}{0}
\setcounter{section}{0}
\renewcommand{\theequation}{\arabic{section}.\arabic{equation}}

\section{Introduction}
\setcounter{equation}{0}

Effects from gravitational physics and quantum physics are expected to be on an equal footing at the Planck scale, which could induce minuscule violations of Lorentz invariance. Such violations were explicitly demonstrated to arise in closed-string theories \cite{Kostelecky:1988zi,Kostelecky:1989jp,Kostelecky:1989jw,Kostelecky:1991ak,Kostelecky:1994rn}, loop quantum gravity \cite{Gambini:1998it,Bojowald:2004bb}, models of noncommutative spacetimes \cite{AmelinoCamelia:1999pm,Carroll:2001ws}, spacetime foam models \cite{Klinkhamer:2003ec,Bernadotte:2006ya,Hossenfelder:2014hha}, (chiral) field theories defined on spacetimes with nontrivial topologies \cite{Klinkhamer:1998fa,Klinkhamer:1999zh,Klinkhamer:2002mj,Ghosh:2017iat}, and Ho\v{r}ava-Lifshitz gravity \cite{Horava:2009uw}. For several decades, violations of this fundamental symmetry have been looked for where until the end of the past millennium, these experimental searches had been rather unsystematic. To be able to compare the results of different experiments to each other and to make theoretical predictions of possible experimental signals, the minimal Standard-Model Extension (SME) was established in \cite{Colladay:1996iz,Colladay:1998fq}. The minimal SME is a field-theory framework parameterizing all power-counting renormalizable Lorentz-violating contributions that are consistent with both coordinate invariance and the gauge structure of the Standard Model. Each contribution is decomposed into controlling coefficients and a field operator suitably contracted to form observer Lorentz scalars. In gravity, Lorentz invariance is a local concept and a parameterization of deviations from local Lorentz invariance, local position invariance, and the weak equivalence principle in terms of minimal operators was delivered in \cite{Kostelecky:2003fs}. The nongravitational part of the SME was extended in \cite{Kostelecky:2009zp,Kostelecky:2011gq,Kostelecky:2013rta} to include all operators of arbitrary mass dimension where this generalization is called the nonminimal SME. It is important to recall that {\em CPT} violation implies Lorentz violation in effective field theory \cite{Greenberg:2002uu}, which is why all {\em CPT}-violating operators are contained in the SME, as well.

Most experimental tests of gravity are performed with classical test bodies, cf.~\cite{Kostelecky:2008ts} for a compilation of all current constraints on controlling coefficients. General Relativity is a classical theory, after all. However, the field-theory description of the SME is not entirely suitable to predict effects of Lorentz violation within classical physics, which is why it would be highly desirable to have a map from the Lagrange density in field theory to the classical Lagrangian of a relativistic, pointlike particle. A map that delivers a classical Lagrangian from the field-theory dispersion equation was constructed explicitly in \cite{Kostelecky:2010hs}. Based on this construction, the first classical Lagrangians for a wide range of controlling coefficients within the minimal SME were obtained in the same paper. These results are exact and comprise Lagrangians for a combination of $a$, $c$, $e$, $f$ coefficients, the full Lagrangian for the $b$ coefficients, and partial results for the $d$, $H$ coefficients. This set of Lagrangians was complemented by results for more involved families of $d$, $g$ coefficients given in~\cite{Colladay:2012rv,Russell:2015gwa}.

Based on some of these findings, the motion of a charged, classical particle under the influence of both a Lorentz-violating background field and an electromagnetic field was studied in \cite{Schreck:2014ama}. In addition, the modified time evolution of a semi-classical analog of spin was examined with the BMT equation. In \cite{Schreck:2016jqn} it was demonstrated that the procedure of finding Lagrangians from the field theory description can be reversed at first order in Lorentz violation. Thus, it is possible to reobtain parts of the SME Hamilton operator from a particular classical Lagrangian by the usual quantization procedure. Last but not least, an approach was developed in \cite{Schreck:2015dsa} to assign Finsler structures to the minimal SME photon sector. This procedure is based on modified refractive indices and the eikonal equation.

The first exact nonminimal Lagrangian was derived in \cite{Schreck:2014hga} for the isotropic operator $\hat{m}^{(5)}$. However, the latter is highly complicated, nontransparent and too unwieldy to be used in applications. Therefore, the point of view changed and subsequently it was found to be more reasonable to obtain classical Lagrangians in the nonminimal SME at first order in Lorentz violation only. Based on the method of Groebner bases, which is a generalization of Gauss' algorithm for nonlinear systems of equations, it was possible to derive Lagrangians for the lowest-dimensional contributions to the spin-degenerate operators, i.e., for $\hat{a}^{(5)}$, $\hat{c}^{(6)}$, $\hat{e}^{(6)}$, $\hat{f}^{(6)}$, and $\hat{m}^{(5)}$ \cite{Schreck:2015seb}. These Lagrangians are natural generalizations of the first-order minimal ones with the controlling coefficients replaced by suitable contractions of the nonminimal coefficients and the four-velocity. An additional scalar parameter before the Lorentz-violating contribution ensures that its mass dimension is consistent with that of the standard term and that it is positively homogeneous of degree one in the velocity.

It was shown in \cite{Kostelecky:2011qz} that such classical Lagrangians can be promoted to Finsler structures by a procedure that has parallels to a Wick rotation. The Lagrangians for the $a$, $c$, $e$, and $f$ coefficients were found to be related to what is known as a Randers structure, whereas the Lagrangian for $b$ was demonstrated to be linked to a hitherto unknown Finsler space that is not of Randers type. Also, the properties of this Finsler space, which is referred to as $b$ space, were investigated in the latter article. Three-dimensional versions of $b$ space were demonstrated to play a role in systems of classical mechanics and magnetostatics \cite{Foster:2015yta}. Besides that, Finsler $b$ space was discovered to have singularities that can be removed with a desingularization procedure \cite{Colladay:2015wra} whose existence is based on the famous Hironaka theorem for algebraic varieties. An alternative to this procedure was presented in the recent work \cite{Colladay:2017bon} related to the formalism of extended Hamiltonians introduced by Dirac. In principle, $b$ space is a special case of a more general type of Finsler structures that are called bipartite and that are focused on in \cite{Kostelecky:2012ac}. Ultimately, the classical first-order Lagrangians of the nonminimal SME obtained in \cite{Schreck:2015seb} were promoted to Finsler structures, as well, by adapting the Wick rotation procedure to the nonminimal SME. The resulting Finsler structures seem to define new kinds of Finsler geometries that had not been studied until then.

The objective of the current paper is to deliver a complete generalization of the results of~\cite{Schreck:2015seb} that covers the whole SME at first order in Lorentz violation. Special emphasis is put on the spin-nondegenerate operators $\hat{b}$, $\hat{d}$, $\hat{H}$, and $\hat{g}$ that were not treated in the previous reference. The paper is organized, as follows. In \secref{sec:classical-lagrangians} we recall the basics on how to obtain a classical Lagrangian from the corresponding field theory based on the SME. The five equations describing the map from the wave packet to a classical, pointlike particle will be introduced. The Lagrangians for the spin-degenerate operators will then be derived and discussed in \secref{sec:results-spin-degenerate-operators} where the results for the spin-nondegenerate operators follow in \secref{sec:results-spin-nondegenerate-operators}. Based on the Lagrangians obtained to that point, in \secref{sec:general-first-order-lagrangian} we will be in a position to state the first-order result covering the full SME, which is the central result of the current article. Finally, all findings will be concluded on in \secref{sec:conclusions}. In addition, it will be demonstrated analytically that the five nonlinear equations are fulfilled at leading order by the classical Lagrangians obtained. As these proofs for the spin-nondegenerate operators are lengthy, they will be relegated to App.~\ref{sec:analytical-proof-b} and \ref{sec:analytical-proof-H}, respectively. Natural units with $\hbar=c=1$ are used unless otherwise stated.

\section{Map from field-theory description to classical Lagrangian}
\label{sec:classical-lagrangians}
\setcounter{equation}{0}

We consider the Lagrange density describing nonminimal Lorentz violation in the fermion sector of the SME. Its explicit form is stated in \cite{Kostelecky:2013rta} and we will recall it as follows:
\begin{subequations}
\label{eq:lagrange-density-fermion-sector}
\begin{align}
\mathcal{L}_{\mathrm{Dirac}}&=\frac{1}{2}\overline{\psi}(\mathrm{i}\cancel{\partial}-m_{\psi}\mathds{1}_4+\hat{\mathcal{Q}})\psi+\text{H.c.}\,, \displaybreak[0]\\[2ex]
\hat{\mathcal{Q}}&=\hat{\mathcal{S}}\mathds{1}_4+\mathrm{i}\hat{\mathcal{P}}\gamma^5+\hat{\mathcal{V}}^{\mu}\gamma_{\mu}+\hat{\mathcal{A}}^{\mu}\gamma^5\gamma_{\mu}+\frac{1}{2}\hat{\mathcal{T}}^{\mu\nu} \sigma_{\mu\nu}\,, \displaybreak[0]\\[2ex]
\hat{\mathcal{S}}&=\hat{e}-\hat{m}\,,\quad\hat{\mathcal{P}}=\hat{f}-\hat{m}_5\,,\quad \hat{\mathcal{V}}^{\mu}=\hat{c}^{\mu}-\hat{a}^{\mu}\,, \displaybreak[0]\\[2ex] \hat{\mathcal{A}}^{\mu}&=\hat{d}^{\mu}-\hat{b}^{\mu}\,,\quad \hat{\mathcal{T}}^{\mu\nu}=\hat{g}^{\mu\nu}-\hat{H}^{\mu\nu}\,.
\end{align}
\end{subequations}
Here, $\psi$ is the Dirac field and $\overline{\psi}\equiv\psi^{\dagger}\gamma^0$ its Dirac conjugate where both fields are defined in Minkowski spacetime with metric $\eta_{\mu\nu}=\mathrm{diag}(1,-1,-1,-1)_{\mu\nu}$. The Dirac matrices $\gamma^{\mu}$ are standard and obey the Clifford algebra $\{\gamma^{\mu},\gamma^{\nu}\}=2\eta^{\mu\nu}\mathds{1}_4$ with the unit matrix $\mathds{1}_4$ in spinor space. Based on the Dirac matrices, we define the chiral matrix $\gamma_5=\gamma^5\equiv \mathrm{i}\gamma^0\gamma^1\gamma^2\gamma^3$ and the commutator $\sigma^{\mu\nu}\equiv (\mathrm{i}/2)[\gamma^{\mu},\gamma^{\nu}]$. Lorentz violation is encoded in the operators $\hat{a}$, $\hat{b}$ $\dots$ $\hat{m}$ where the fermion mass is called $m_{\psi}$ to avoid confusions with $\hat{m}$. According to the properties of the Lorentz-violating operators under proper, orthochronous Lorentz transformations as well as parity and time reversal transformations, they are grouped into scalars $\hat{\mathcal{S}}$, pseudo-scalars $\hat{\mathcal{P}}$, vectors $\hat{\mathcal{V}}^{\mu}$, axialvectors $\hat{\mathcal{A}}^{\mu}$, and two-tensors $\hat{\mathcal{T}}^{\mu\nu}$. It has been well-known for some time that $\hat{m}_5$ can be removed by a chiral transformation of the Dirac field \cite{Kostelecky:2013rta}. However, this operator will be kept for completeness.

The map from the field-theory Lagrange density to the classical Lagrange function of a relativistic, pointlike particle of mass $m_{\psi}$ moving with four-velocity $u^{\mu}$ is governed by the following set of five ordinary nonlinear equations:
\begin{subequations}
\label{eq:set-equations-lagrangians}
\begin{align}
\label{eq:dispersion-relation}
\mathcal{R}(p)&=0\,, \displaybreak[0]\\[2ex]
\label{eq:group-velocity-correspondence}
\frac{\partial p_0}{\partial p_i}&=-\frac{u^i}{u^0}\,,\quad i\in\{1,2,3\}\,, \displaybreak[0]\\[2ex]
\label{eq:euler-equation}
L&=-p_{\mu}u^{\mu}\,,\quad p_{\mu}=-\frac{\partial L}{\partial u^{\mu}}\,,
\end{align}
\end{subequations}
where $p_{\mu}=\mathrm{i}\partial_{\mu}$ is the momentum that appears in the Fourier decomposition of a wave packet into plane waves. The first equation depends on the momentum only and it is simply the dispersion equation following from the determinant of the modified Dirac operator. The set of three equations~(\ref{eq:group-velocity-correspondence}) identifies the group velocity of the centroid of a quantum wave packet with the three-velocity $\mathbf{u}/u^0$ of the classical particle. The convention is such that the four-momentum components have lower indices whereas the four-velocity components are supposed to have upper ones. Hence, an additional sign must be taken into account on the right-hand side. Last but not least, the classical Lagrange function is supposed to be positively homogeneous in the velocity, which means that $L(\lambda u)=\lambda L(u)$ for $\lambda>0$. This property grants parameterization invariance of the action, which is a must-have in physics, as the action should only depend on the path, but not on the way how it is parameterized. The fifth of the equations above is famous Euler's theorem that follows from exactly the latter characteristic of the Lagrangian. The canonical momentum is defined with a minus sign to grant the correct sign for the nonrelativistic kinetic energy.

The five equations depend on the four-momentum components, the four-velocity components, and the classical Lagrangian. Solving them for $p^{\mu}=p^{\mu}(u)$ and $L=L(u)$ corresponds to an explicit construction of the map. However, as the system is nonlinear, it is highly nontrivial to solve it --- even in the context of the minimal SME. Groebner bases deliver a tool to treat the equations systematically, which has proven extremely useful to derive Lagrangians in the nonminimal SME~\cite{Schreck:2015seb}. In what follows, classical Lagrangians will be obtained that fulfill Eqs.~(\ref{eq:set-equations-lagrangians}) at leading order in Lorentz violation. The complete family of nonminimal coefficients ought to be covered. We will restrict our analysis to particles. Classical Lagrangians corresponding to antiparticle dispersion relations can simply be obtained from the particle Lagrangians via the substitution $m_{\psi}\mapsto -m_{\psi}$, cf.~\cite{Kostelecky:2010hs,Colladay:2012rv}.

\section{Results for spin-degenerate operators}
\label{sec:results-spin-degenerate-operators}

Let us start deriving the classical Lagrangians for the spin-degenerate operators $\hat{a}$, $\hat{c}$, $\hat{e}$, $\hat{f}$, and $\hat{m}$. Our approach shall be exemplified by the $a$ coefficients. The minimal $a$ coefficients are comprised by an observer four-vector and they are contracted with a field operator of mass dimension 3. In the nonminimal SME, the number of additional derivatives in the field operator subsequently increases by two whereupon the number of indices of the controlling coefficients rises by two and the mass dimension decreases by two. Hence, the operator $\hat{a}$ contains the coefficients $a^{(3)\mu}$, $a^{(5)\mu\nu\varrho}$, $a^{(7)\mu\nu\varrho\sigma\lambda}$, etc. The Lagrangian for $d=3$ was obtained in \cite{Kostelecky:2010hs} where for $d=5$ it was found in \cite{Schreck:2015seb} based on the method of Groebner bases.
We recall both Lagrangians:
\begin{subequations}
\label{eq:lagrangians-a-dimension-3-and-5}
\begin{align}
L^{a^{(3)}}&=-m_{\psi}\sqrt{u^2}-a_{\ast}^{(3)}\,,\quad a_{\ast}^{(3)}\equiv a_{\mu}^{(3)}u^{\mu}\,, \displaybreak[0]\\[2ex]
L^{\hat{a}^{(5)}}&=-m_{\psi}\sqrt{u^2}-\frac{m_{\psi}^2}{u^2}\hat{a}_{\ast}^{(5)}+\dots\,,\quad \hat{a}_{\ast}^{(5)}\equiv a^{(5)}_{\mu\nu\varrho}u^{\mu}u^{\nu}u^{\varrho}\,.
\end{align}
\end{subequations}
The minimal result is exact and the nonminimal one is valid at first order in Lorentz violation. Neglected higher-order contributions will be indicated as ellipses. Due to observer Lorentz invariance, consistency of the mass dimension, and positive homogeneity of degree 1 in the velocity, the perturbative form of the Lagrangian is quite restricted. The Lorentz-violating contribution must involve the component coefficients and suitable observer Lorentz scalars $\hat{a}_{\ast}^{(d)}$ can only be formed by contracting the latter with the only four-vector available, which is the four-velocity.\footnote{Throughout the article, an asterisk indicates controlling coefficients suitably contracted with four-velocities. This notation was introduced in \cite{Schreck:2015seb} and it will be adopted for consistency.} The mass dimension is made consistent with that of the standard term by introducing additional powers of the fermion mass. Positive homogeneity of the second term is restored by suitable powers of the Lorentz scalar $u^2$ in the denominator. These general arguments enable us to propose an \textit{Ansatz} for arbitrary odd $d\geq 3$:
\begin{subequations}
\label{eq:lagrangian-a-ansatz}
\begin{align}
L^{\hat{a}^{(d)}}&=-m_{\psi}\sqrt{u^2}-\hat{A}_{\ast}+\dots\,, \\[2ex]
\hat{A}_{\ast}&\equiv \Xi_a\hat{a}_{\ast}\,,\quad \Xi_a=\Xi_a^{(d)}=\frac{m_{\psi}^{d-3}}{(u^2)^{(d-3)/2}}\,, \\[2ex]
\hat{a}_{\ast}&=\hat{a}_{\ast}^{(d)}\equiv a^{(d)}_{\mu\alpha_1\dots\alpha_{d-3}}u^{\mu}u^{\alpha_1}\dots u^{\alpha_{d-3}}\,,
\end{align}
\end{subequations}
where this result is supposed to be valid at first order in the controlling coefficients. It is convenient to define a second Lorentz scalar $\hat{A}_{\ast}$ that corresponds to the product of the consistency factor $\Xi_a$ and the Lorentz scalar obtained by contracting the controlling coefficients with the four-velocity. Note that $\hat{A}_{\ast}$ is homogeneous of degree 1 in contrast to $\hat{a}_{\ast}$, as each of the $d-2$ indices is contracted with a four-velocity to be compensated by the factor $(u^2)^{(d-3)/2}$ in the denominator. The usefulness of $\hat{A}_{\ast}$ will become clear below when a general proof of the validity of the Lagrangian at leading order in Lorentz violation will be delivered.

It can be checked that for $d=3$, 5 the \textit{Ansatz} reproduces the already known results of \eqref{eq:lagrangians-a-dimension-3-and-5}. Note that in what follows, the index indicating the mass dimension at various quantities will be dropped to simplify the notation, i.e., it will be mentioned only when necessary. As a starting point, we were interested in testing whether this \textit{Ansatz} could be valid for higher-dimensional operators such as that for $d=7$. The large number of nonzero controlling coefficients makes an analytical check challenging. Hence, we used \textit{Mathematica} to demonstrate numerically that \eqref{eq:lagrangian-a-ansatz} fulfills Eqs.~(\ref{eq:set-equations-lagrangians}) at first order in the controlling coefficients.

Subsequently, we discovered that a formal proof is feasible and it is demonstrated as follows, i.e., we will show that the Lagrangian fulfills the defining equations at leading order in Lorentz violation. The starting point is the covariant momentum as a function of the four-velocity, which is obtained from the Lagrangian via
\begin{equation}
\label{eq:momentum-from-lagrangian}
p_{\mu}=-\frac{\partial L^{\hat{a}^{(d)}}}{\partial u^{\mu}}=m_{\psi}\frac{u_{\mu}}{\sqrt{u^2}}+\frac{\partial\hat{A}_{\ast}}{\partial u^{\mu}}\,,
\end{equation}
cf.~\eqref{eq:euler-equation}. Note that here it is already much more convenient to express the Lagrangian in terms of $\hat{A}_{\ast}$, as it is not necessary to include the derivative of $\Xi_a$ explicitly. When the Lagrangian is correct, as it stands, this momentum must obey the dispersion equation at first order in Lorentz violation, which is quite simple to show for the $a$ coefficients:
\begin{subequations}
\begin{align}
\mathcal{R}&=(p-\hat{a})^2-m_{\psi}^2=p^2-2\hat{a}\cdot p-m_{\psi}^2+\dots\,, \\[2ex]
\hat{a}_{\mu}&=\hat{a}^{(d)}_{\mu}\equiv a_{\mu\alpha_1\dots\alpha_{d-3}}p^{\alpha_1}\dots p^{\alpha_{d-3}}\,.
\end{align}
\end{subequations}
Here $\hat{a}_{\mu}$ is a four-vector formed from a suitable contraction of the controlling coefficients with the four-momentum. At leading order in Lorentz violation, it suffices to replace all four-momenta by the standard expression $m_{\psi}u_{\mu}/\sqrt{u^2}$ producing $\hat{a}_{\mu}\approx \Xi_a \hat{a}_{\ast\mu}$. Inserting the momentum and the latter approximate relationship into the dispersion relation gives
\begin{align}
\mathcal{R}&=m_{\psi}^2+\frac{2m_{\psi}}{\sqrt{u^2}}u^{\mu}\frac{\partial \hat{A}_{\ast}}{\partial u^{\mu}}-\frac{2m_{\psi}}{\sqrt{u^2}}\hat{A}_{\ast}-m_{\psi}^2+\dots=0\,.
\end{align}
It is very convenient to employ Euler's theorem, which is applicable, as $\hat{A}_{\ast}$ is positively homogeneous of degree 1 in the four-velocity:
\begin{equation}
u^{\mu}\frac{\partial\hat{A}_{\ast}}{\partial u^{\mu}}=\hat{A}_{\ast}\,.
\end{equation}
Hence, the dispersion equation is fulfilled when neglecting higher-order contributions in Lorentz violation. The next step is to verify \eqref{eq:group-velocity-correspondence}. We do this by computing the first implicit derivative of the dispersion equation with respect to $p_i$, replace $\partial p_0/\partial p_i$ by $-u^i/u^0$, and insert the canonical momentum of \eqref{eq:momentum-from-lagrangian}:
\begin{align}
\label{eq:implicit-derivative-dispersion-equation-a}
\frac{\partial\mathcal{R}}{\partial p_i}&=2p_0\frac{\partial p_0}{\partial p_i}+2p^i-2\frac{\partial(\hat{a}\cdot p)}{\partial p_i}+\dots \notag \displaybreak[0]\\
&=2\left[m_{\psi}\frac{u^0}{\sqrt{u^2}}+\frac{\partial \hat{A}_{\ast}}{\partial u^0}\right]\left(-\frac{u^i}{u^0}\right)+2\left(m_{\psi}\frac{u^i}{\sqrt{u^2}}-\frac{\partial \hat{A}_{\ast}}{\partial u^i}\right)-2\frac{\partial(\hat{a}\cdot p)}{\partial p_i} \notag \displaybreak[0]\\
&=-2\left(\frac{\partial \hat{A}_{\ast}}{\partial u^0}\frac{u^i}{u^0}+\frac{\partial \hat{A}_{\ast}}{\partial u^i}\right)-2\frac{\partial(\hat{a}\cdot p)}{\partial p_i}=0\,,
\end{align}
where the derivative of $\hat{a}\cdot p$ with respect to the momentum was expressed in terms of the derivative of $\hat{A}_{\ast}$ with respect to the four-velocity components:
\begin{align}
\label{eq:derivative-lv-operator}
\frac{\partial(\hat{a}\cdot p)}{\partial p_i}&=m_{\psi}\frac{\partial(\Xi_a\hat{a}_{\ast\mu}u^{\mu}/\sqrt{u^2})}{\partial u^{\sigma}}\frac{\partial u^{\sigma}}{\partial p_i}+\dots=m_{\psi}\left(\frac{\partial(\hat{A}_{\ast}/\sqrt{u^2})}{\partial u^0}\frac{\partial u^0}{\partial p_i}+\frac{\partial(\hat{A}_{\ast}/\sqrt{u^2})}{\partial u^j}\frac{\partial u^j}{\partial p_i}\right) \notag \\
&=-\left(\frac{\partial \hat{A}_{\ast}}{\partial u^0}\frac{u^i}{u^0}+\frac{\partial \hat{A}_{\ast}}{\partial u^i}\right)\,.
\end{align}
To do so, the leading-order correspondence between four-velocity and four-momentum and the related derivatives are needed
\begin{subequations}
\label{eq:derivatives-four-velocity}
\begin{align}
\frac{u^{\mu}}{\sqrt{u^2}}&=\frac{p^{\mu}}{\sqrt{p^2}}=\frac{p^{\mu}}{m_{\psi}}\,, \\[2ex]
\frac{1}{\sqrt{u^2}}\frac{\partial u^0}{\partial p_i}&=\frac{1}{m_{\psi}}\frac{p_i}{p^0}=-\frac{1}{m_{\psi}}\frac{u^i}{u^0}\,,\quad \frac{1}{\sqrt{u^2}}\frac{\partial u^j}{\partial p_i}=-\frac{1}{m_{\psi}}\delta^{ij}\,,
\end{align}
\end{subequations}
in combination with the following property for a generic function $f=f(u^2)$ that depends on the four-velocity squared only:
\begin{equation}
\label{eq:derivative-function-f-u2}
\frac{u^i}{u^0}\frac{\partial f(u^2)}{\partial u^0}+\frac{\partial f(u^2)}{\partial u^i}=\frac{u^i}{u^0}\frac{\partial f}{\partial(u^2)}2u^0-\frac{\partial f}{\partial(u^2)}2u^i=0\,.
\end{equation}
Thus, the term involving the derivative of $1/\sqrt{u^2}$ vanishes in \eqref{eq:derivative-lv-operator} whereupon the remaining result compensates the first term in \eqref{eq:implicit-derivative-dispersion-equation-a}. As the Lagrangian is positively homogeneous of degree 1 in the velocity, \eqref{eq:euler-equation} is fulfilled automatically.

To summarize, in contrast to the derivations of classical Lagrangians in \cite{Schreck:2015seb}, which relied on the method of Groebner bases, the current Lagrangian was simply obtained by making a suitable guess that is in accordance with observer Lorentz invariance and positive homogeneity of first degree in the velocity. Furthermore, it ought to reproduce the already known results. The same procedure is now successfully employed to arrive at the classical Lagrangian for the remaining spin-degenerate operators. The general result for the classical Lagrangian at leading order in Lorentz violation is reasonably expressed in the form
\begin{equation}
\label{eq:lagrangian-spin-degenerate}
L^{\hat{x}^{(d)}}=-m_{\psi}\sqrt{u^2}-\hat{X}_{\ast}+\dots\,,\quad \hat{X}_{\ast}=\Xi_x\hat{x}_{\ast}\,.
\end{equation}
\begin{table}
\centering
\begin{tabular}{cccrlrlccc}
\toprule
$\hat{x}_{\ast}$ & Explicit contraction & $\Xi_x$ & \multicolumn{2}{c}{Radial derivative} & \multicolumn{2}{c}{Correspondence} \\
\colrule
$\hat{a}_{\ast}$ & $a^{(d)}_{\mu\alpha_1\dots \alpha_{d-3}}u^{\mu}u^{\alpha_1}\dots u^{\alpha_{d-3}}$ & $m_{\psi}^{d-3}/(u^2)^{(d-3)/2}$ & $u^{\mu}\frac{\partial \hat{a}_{\ast}}{\partial u^{\mu}}$ & $=(d-2)\hat{a}_{\ast}$ & $\hat{a}_{\mu}$ & $\approx \Xi_a \hat{a}_{\ast\mu}$ \\
$\hat{c}_{\ast}$ & $c^{(d)}_{\mu\alpha_1\dots \alpha_{d-3}}u^{\mu}u^{\alpha_1}\dots u^{\alpha_{d-3}}$ & $-m_{\psi}^{d-3}/(u^2)^{(d-3)/2}$ & $u^{\mu}\frac{\partial \hat{c}_{\ast}}{\partial u^{\mu}}$ & $=(d-2)\hat{c}_{\ast}$ & $\hat{c}_{\mu}$ & $\approx -\Xi_c \hat{c}_{\ast\mu}$ \\
$\hat{e}_{\ast}$ & $e^{(d)}_{\alpha_1\dots \alpha_{d-3}}u^{\alpha_1}\dots u^{\alpha_{d-3}}$ & $-m_{\psi}^{d-3}/(u^2)^{(d-4)/2}$ & $u^{\mu}\frac{\partial \hat{e}_{\ast}}{\partial u^{\mu}}$ & $=(d-3)\hat{e}_{\ast}$ & $\hat{e}_{\mu}$ & $\approx-\frac{\Xi_e}{m_{\psi}}\hat{e}_{\ast\mu}$ \\
$\hat{f}_{\ast}$ & $f^{(d)}_{\alpha_1\dots \alpha_{d-3}}u^{\alpha_1}\dots u^{\alpha_{d-3}}$ & $m_{\psi}^{2d-7}/[2(u^2)^{(2d-7)/2}]$ & $u^{\mu}\frac{\partial \hat{f}_{\ast}^2}{\partial u^{\mu}}$ & $=2(d-3)\hat{f}_{\ast}^2$ & $\hat{f}_{\mu}\hat{f}_{\nu}$ & $\approx\frac{2\sqrt{u^2}\Xi_f}{m_{\psi}}\hat{f}_{\ast\mu}\hat{f}_{\ast\nu}$ \\
$\hat{m}_{\ast}$ & $m^{(d)}_{\alpha_1\dots \alpha_{d-3}}u^{\alpha_1}\dots u^{\alpha_{d-3}}$ & $m_{\psi}^{d-3}/(u^2)^{(d-4)/2}$ & $u^{\mu}\frac{\partial \hat{m}_{\ast}}{\partial u^{\mu}}$ & $=(d-3)\hat{m}_{\ast}$ & $\hat{m}$ & $\approx\frac{\Xi_m}{\sqrt{u^2}}\hat{m}_{\ast}$ \\
$\hat{\mathcal{S}}_{\ast}$ & $\mathcal{S}^{(d)}_{\alpha_1\dots\alpha_{d-3}}u^{\alpha_1}\dots u^{\alpha_{d-3}}$ & $-m_{\psi}^{d-3}/(u^2)^{(d-4)/2}$ & $u^{\mu}\frac{\partial\hat{\mathcal{S}}_{\ast}}{\partial u^{\mu}}$ & $=(d-3)\hat{\mathcal{S}}_{\ast}$ & $\hat{\mathcal{S}}$ & $\approx-\frac{\Xi_{\mathcal{S}}}{\sqrt{u^2}}\hat{\mathcal{S}}_{\ast}$ \\
$\hat{\mathcal{V}}_{\ast}$ & $\mathcal{V}^{(d)}_{\mu\alpha_1\dots\alpha_{d-3}}u^{\mu}u^{\alpha_1}\dots u^{\alpha_{d-3}}$ & $-m_{\psi}^{d-3}/(u^2)^{(d-3)/2}$ & $u^{\mu}\frac{\partial\hat{\mathcal{V}}_{\ast}}{\partial u^{\mu}}$ & $=(d-2)\hat{\mathcal{V}}_{\ast}$ & $\hat{\mathcal{V}}_{\mu}$ & $\approx-\Xi_{\mathcal{V}}\hat{\mathcal{V}}_{\ast\mu}$ \\
\botrule
\end{tabular}
\caption{Parameters of the generic classical Lagrangian of \eqref{eq:lagrangian-spin-degenerate}. The first column states the observer scalar employed in the Lagrangian where the entries in the second column give the corresponding explicit expressions. The consistency factors ensuring the correct mass dimension and positive homogeneity of the Lorentz-violating contribution are listed in the third column. The fourth column shows Euler's theorem for each of the Lorentz scalars defined in the first two columns. Last but not least, the fifth column states the leading-order correspondences between the Lorentz-violating operators transformed to momentum space and the quantities associated with each four-momentum replaced by the four-velocity.}
\label{tab:parameters-spin-degenerate-lagrangians}
\end{table}
Here $\hat{x}_{\ast}=\hat{x}_{\ast}^{(d)}$ is a total contraction of controlling coefficients with an appropriate combination of four-velocities. The parameter $\Xi_x=\Xi_x^{(d)}$ only involves the particle mass and a suitable power of the observer scalar $u^2$. In the general case, it is also convenient to define quantities $\hat{X}_{\ast}=\hat{X}^{(d)}_{\ast}$ that are positively homogeneous of degree 1 in the velocity. The explicit expressions of these quantities for the complete set of spin-degenerate operators are listed in \tabref{tab:parameters-spin-degenerate-lagrangians}.

Several remarks on the classical Lagrangian of \eqref{eq:lagrangian-spin-degenerate} are in order. First, the Lagrangian is a function of the particle mass, the four-velocity, and the controlling coefficients. It is a sum of the standard term $L=-m_{\psi}\sqrt{u^2}$ and a contribution that is linear in the controlling coefficients, i.e., it reduces to the standard result for vanishing Lorentz violation. Second, the Lagrangian is formed from observer Lorentz scalars to render it a Lorentz scalar, as expected. Third, it has to be of mass dimension 1 where additional powers of masses must be introduced in the Lorentz-violating term such that its mass dimension corresponds to the mass dimension of the standard term. Fourth, additional powers of the four-velocity squared are needed to make the nonstandard contribution homogeneous of degree 1 in the velocity. Fifth, there is the correspondence $\hat{a}_{\ast}\leftrightarrow -m_{\psi}\hat{e}_{\ast}$, which is the generalization of $a_{\mu}\leftrightarrow -m_{\psi}e_{\mu}$ found for the minimal SME \cite{Russell:2015gwa}. Sixth, the dimensionless number $\hat{X}_{\ast}/m_{\psi}$ must be $\ll 1$ such that the first-order approximation is justified. This requirement translates into the additional condition that $\mathbf{u}^2$ should not lie in the close vicinity of $u_0^2$. When we use a parameterization of the particle trajectory such that $u^0=1$ and $\mathbf{u}=\mathbf{v}$ with the three-velocity $\mathbf{v}$, this condition means that the particle should not travel with a velocity too close to the speed of light. The analog condition in momentum space is that the energy and momentum must be small enough, as the relevance of a nonminimal contribution rises with the momentum.

For the remaining spin-degenerate operators, the proof that \eqref{eq:lagrangian-spin-degenerate} fulfills Eqs.~(\ref{eq:set-equations-lagrangians}) works completely analogous when the corresponding expressions of \tabref{tab:parameters-spin-degenerate-lagrangians} are employed. Furthermore, it is now quite convenient to generalize the Lagrangian of \eqref{eq:lagrangian-spin-degenerate} to the situation when operators of different mass dimensions are added, e.g., $\hat{a}_{\mu}\mapsto \sum_{d\geq 3} a^{(d)}_{\mu\alpha_1\dots \alpha_{d-3}}p^{\alpha_1}\dots p^{\alpha_{d-3}}$ for the $a$ coefficients. We then simply have to replace the above observer scalar $\hat{X}_{\ast}^{(d)}$ for a particular mass dimension $d$ by a sum, i.e., $\hat{X}_{\ast}^{(d)}\mapsto\sum_d \hat{X}_{\ast}^{(d)}$ where $d$ runs over suitable values permitted for the operator $\hat{x}^{(d)}$. As each individual summand of the introduced sum is positively homogeneous of degree 1 in the velocity, the sum itself will have that property, too. Hence, the proof of the validity of the Lagrangian can be taken over completely. Last but not least, the expansion of Lorentz-violating operators in terms of momenta leads to the additional requirement $\hat{X}^{(d+2)}_{\ast}\ll\hat{X}^{(d)}_{\ast}$ such that each contribution is suppressed compared to the previous one.

\subsection{Effective coefficients}

At first order in Lorentz violation, suitable field redefinitions allow for combining controlling coefficients of different mass dimension such that new coefficients can be defined that are known as effective, cf.~Eqs.~(27) of \cite{Kostelecky:2013rta}. Based on this observation, it is possible to define effective observer scalars $\hat{a}_{\ast,\mathrm{eff}}$, $\hat{c}_{\ast,\mathrm{eff}}$. Generically,
\begin{equation}
\hat{X}^{(d)}_{\ast,\mathrm{eff}}\equiv\Xi_x^{(d)}\hat{x}^{(d)}_{\ast,\mathrm{eff}}\,,\quad\hat{x}^{(d)}_{\ast,\mathrm{eff}}\equiv x^{(d)}_{\mathrm{eff},\mu\alpha_1\dots\alpha_{d-3}}u^{\mu}u^{\alpha_1}\dots u^{\alpha_{d-3}}\,.
\end{equation}
Hence, in the classical description, each momentum must simply be replaced by the four-velocity. Now, the effective observer scalar linked to the operator $\hat{a}$ can be expressed via $\hat{A}_{\ast}$ and $\hat{E}_{\ast}$ of different dimensionalities:
\begin{align}
\hat{A}^{(d)}_{\ast,\mathrm{eff}}&\equiv\Xi_a^{(d)}a^{(d)}_{\mathrm{eff},\mu\alpha_1\dots\alpha_{d-3}}u^{\mu}u^{\alpha_1}\dots u^{\alpha_{d-3}} \notag \displaybreak[0]\\
&=\Xi_a^{(d)}\left(a^{(d)}_{\mu\alpha_1\dots\alpha_{d-3}}-\frac{1}{m_{\psi}}\eta_{\mu\alpha_1}e^{(d-1)}_{\alpha_2\dots\alpha_{d-3}}\right)u^{\mu}u^{\alpha_1}\dots u^{\alpha_{d-3}} \notag \displaybreak[0]\\
&=\Xi_a^{(d)}\hat{a}_{\ast}^{(d)}-\frac{m_{\psi}^{d-4}}{(u^2)^{(d-5)/2}}e^{(d-1)}_{\alpha_2\dots\alpha_{d-3}}u^{\alpha_2}\dots u^{\alpha_{d-3}} \notag \displaybreak[0]\\
&=\Xi_a^{(d)}\hat{a}_{\ast}^{(d)}-\Xi_e^{(d-1)}\hat{e}^{(d-1)}_{\ast}=\hat{A}_{\ast}^{(d)}-\hat{E}_{\ast}^{(d-1)}\,,
\end{align}
which is consistent with the parameters listed in \tabref{tab:parameters-spin-degenerate-lagrangians}. An analog correspondence can be derived for the effective observer scalar related to $\hat{c}$:
\begin{equation}
\hat{C}^{(d)}_{\ast,\mathrm{eff}}=\hat{C}_{\ast}^{(d)}-\hat{M}^{(d-1)}_{\ast}\,.
\end{equation}
Because of these connections, it is possible to associate Lagrangians to effective coefficients and at first order in Lorentz violation they are given by
\begin{subequations}
\begin{align}
L^{\hat{a}^{(d)}_{\mathrm{eff}}}&=-m_{\psi}\sqrt{u^2}-\hat{A}^{(d)}_{\ast}+\hat{E}^{(d-1)}_{\ast}\,, \displaybreak[0]\\[2ex]
L^{\hat{c}^{(d)}_{\mathrm{eff}}}&=-m_{\psi}\sqrt{u^2}-\hat{C}^{(d)}_{\ast}+\hat{M}^{(d-1)}_{\ast}\,.
\end{align}
\end{subequations}
These results already demonstrate how at leading order in Lorentz violation, Lagrangians for different component coefficients can be composed to obtain new results for combinations of coefficients.

\subsection{Map between vector $c$ and pseudoscalar $f$ coefficients}

It is well-known that the minimal $f$ coefficients can be mapped onto the $c$ coefficients by a spinor redefinition \cite{Altschul:2006ts}. The structure of the map is such that it involves only bilinear combinations of $f$ coefficients and at leading order, it is given by $c^{(4)}_{\mu\nu}\approx -f^{(4)}_{\mu}f^{(4)}_{\nu}/2$. Comparing the Lagrangians for the $c$ and $f$ coefficients with each other, reveals the following correspondence at leading order:
\begin{subequations}
\label{eq:correspondence-c-f}
\begin{align}
c_{\ast}^{(d)}&\leftrightarrow -\frac{m_{\psi}^{d-4}}{2(u^2)^{(d-4)/2}}(f_{\ast}^{(d)})^2\,, \displaybreak[0]\\[2ex]
c_{\ast\mu\nu}^{(d)}&\leftrightarrow -\frac{m_{\psi}^{d-4}}{2(u^2)^{(d-4)/2}}f_{\ast\mu}^{(d)}f_{\ast\nu}^{(d)}\,.
\end{align}
\end{subequations}
In the second correspondence, the tensor structure on both sides has been extracted. At leading order, the part of the $c$ coefficients that is symmetric in the first two indices contributes to the Lagrangian only. The known map within the minimal SME is reproduced for $d=4$. However, note that Eqs.~(\ref{eq:correspondence-c-f}) do not provide a direct map between the controlling coefficients, but just between certain contractions of the $c$ and $f$ coefficients with the four-velocity. Indeed, there exists the alternative possibility of mapping certain controlling coefficients directly to each other according to the following rule:
\begin{equation}
\label{eq:map-between-c-f-coefficients}
c_{\mu\nu\alpha_2\dots\alpha_{d-3}}^{(d)}\leftrightarrow -\frac{1}{2}f^{(d/2+2)}_{\mu\alpha_2\dots\alpha_{d/2-1}}f^{(d/2+2)}_{\nu\alpha_{d/2+1}\dots\alpha_{d-2}}\,.
\end{equation}
It can be deduced immediately that this generalization contains the correspondence within the minimal SME. Furthermore, counting the numbers of indices on each side produces $d-2$, i.e., the map is consistent in this respect. We obtain the Lagrangian for $\hat{f}^{(d)}$ by inserting the latter map into the Lagrangian for $\hat{c}^{(d)}$ and adapting the mass dimension appropriately:
\begin{align}
L^{\hat{c}^{(d)}}&=-m_{\psi}\sqrt{u^2}-\Xi_c^{(d)}\hat{c}_{\ast}^{(d)}\leftrightarrow -m_{\psi}\sqrt{u^2}+\frac{\Xi_c^{(d)}}{2}(\hat{f}_{\ast}^{(d/2+2)})^2 \notag \\
&=-m_{\psi}\sqrt{u^2}+\frac{\Xi_c^{(2d'-4)}}{2}(\hat{f}_{\ast}^{(d')})^2=-m_{\psi}\sqrt{u^2}+\Xi_f^{(d')}(\hat{f}_{\ast}^{(d')})^2=L^{\hat{f}^{(d')}}\,.
\end{align}
There is one caveat, though. In principle, $d=4+2n$ with $n\in\mathbb{N}_0$ can be chosen for the $c$ coefficients, which directly produces a product of two $f^{(4+n)}_{\mu\alpha_2\dots\alpha_{n+1}}$ on the right-hand side of \eqref{eq:map-between-c-f-coefficients}. This counting would permit arbitrary mass dimensions for $f$ as long as they are $\geq 4$. In analogy to the $c$ coefficients, the mass dimension of the $f$ coefficients only takes values $d=4+2n$, i.e., there is a contradiction. The latter is resolved when restricting the map to $c^{(4n+4)}_{\mu\nu\alpha_2\dots\alpha_{4n+1}}$. For $n=1$ this means that $f^{(6)}_{\mu\alpha_1\alpha_2}$ can only be mapped to $c^{(8)}_{\mu\nu\alpha_2\dots\alpha_5}$ where the case $n=2$ describes a mapping between $f_{\mu\alpha_1\dots\alpha_4}^{(8)}$ and $c_{\mu\nu\alpha_2\dots\alpha_9}^{(12)}$, etc. Therefore, only a subset of $c$ coefficients has a direct connection to $f$ coefficients, whereas the $c_{\mu\nu\alpha_2\alpha_3}^{(6)}$, $c_{\mu\nu\alpha_2\dots\alpha_7}^{(10)}$, etc. do not have an $f$ counterpart.

\section{Results for spin-nondegenerate operators}
\label{sec:results-spin-nondegenerate-operators}
\setcounter{equation}{0}

The base for obtaining classical Lagrangians for the spin-degenerate operators was laid in \cite{Schreck:2015seb}, whereas spin-nondegenerate operators were not considered in the latter paper. Hence, not a single classical Lagrangian has been derived for the nonminimal operators $\hat{b}$, $\hat{d}$, $\hat{H}$, and $\hat{g}$ until now. However, at least a couple of minimal results are known such as that for the $b$ coefficients, cf.~Eq.~(12) of \cite{Kostelecky:2010hs} for $a^{(3)\mu}=0$:
\begin{equation}
L^{b^{(3)}}=-m_{\psi}\sqrt{u^2}\mp\sqrt{(b^{(3)}\cdot u)^2-(b^{(3)})^2u^2}\,,
\end{equation}
where there are two distinct Lagrangians for particles because of the spin-nondegeneracy. When we assume that the previously used technique works for the spin-nondegenerate operators, as well, we could propose a proper \textit{Ansatz} that is in accordance with observer Lorentz invariance, positive homogeneity of degree 1 in the velocity, and with the minimal result $L^{b^{(3)}}$:
\begin{subequations}
\label{eq:classical-lagrangian-operator-b}
\begin{align}
L^{\hat{b}^{(d)}}&=-m_{\psi}\sqrt{u^2}\mp \hat{\mathscr{B}}_{\ast}+\dots\,,\quad \hat{\mathscr{B}}_{\ast}=\sqrt{\hat{B}_{\ast}^2-(\hat{B}_{\ast}^{\mu}\hat{B}_{\ast\mu})u^2}\,, \displaybreak[0]\\[2ex]
\hat{B}_{\ast\mu}&\equiv \Xi_b\hat{b}_{\ast\mu}\,,\quad \hat{B}_{\ast}\equiv \Xi_b\hat{b}_{\ast}\,,\quad \Xi_b=\Xi_b^{(d)}=\frac{m_{\psi}^{d-3}}{(u^2)^{(d-3)/2}}\,, \displaybreak[0]\\[2ex]
\hat{b}_{\ast\mu}&=\hat{b}^{(d)}_{\ast\mu}\equiv b_{\mu\alpha_1\dots\alpha_{d-3}}^{(d)}u^{\alpha_1}\dots u^{\alpha_{d-3}}\,,\quad \hat{b}_{\ast}=\hat{b}_{\ast}^{(d)}\equiv \hat{b}_{\ast\mu}u^{\mu}\,.
\end{align}
\end{subequations}
It was checked numerically for the dimension-5 $b$ coefficients that this guess obeys Eqs.~(\ref{eq:set-equations-lagrangians}). An analytical proof for arbitrary odd $\geq 3$ is tedious but feasible. Readers who are only interested in the results can skip the proof, which is why it has been moved to \appref{sec:analytical-proof-b}. In this context it is crucial to recall that at the level of the dispersion equation, the $b$ and $d$ coefficients contribute to the pseudovector operator $\hat{\mathcal{A}}^{\mu}$, cf. the Lagrange density of \eqref{eq:lagrange-density-fermion-sector}. Hence, the dispersion equation at leading order in Lorentz violation follows from the dispersion equation of $\hat{b}^{\mu}$ in replacing the latter by $-\hat{d}^{\mu\nu}\overline{p}_{\nu}$ where $\overline{p}_{\mu}\equiv (E_0,\mathbf{p})_{\mu}$ with the standard dispersion relation $E_0=(\mathbf{p}^2+m_{\psi}^2)^{1/2}$. The correspondence at the level of classical Lagrangians in then simply $\hat{b}_{\ast\mu}\leftrightarrow -\hat{d}_{\ast\mu\nu}u^{\nu}=-\hat{d}_{\ast\mu}$ at first order in Lorentz violation. The analytical proof for the $b$ coefficients can literally be taken over to the $d$ coefficients where the only difference is that the mass dimension takes even values $\geq 4$.
\begin{table}
\centering
\subfloat[]{\label{tab:parameters-spin-nondegenerate-lagrangians}\begin{tabular}{cccrlrlcc}
\toprule
$\hat{x}_{\ast},\hat{x}_{\ast\mu}$ & Explicit contraction & $\Xi_x$ & \multicolumn{2}{c}{Radial derivative} & \multicolumn{2}{c}{Correspondence} \\
\colrule
$\hat{b}_{\ast}$ & $b^{(d)}_{\mu\alpha_1\dots \alpha_{d-3}}u^{\mu}u^{\alpha_1}\dots u^{\alpha_{d-3}}$ & $m_{\psi}^{d-3}/(u^2)^{(d-3)/2}$ & $u^{\mu}\frac{\partial \hat{b}_{\ast}}{\partial u^{\mu}}$ & $=(d-2)\hat{b}_{\ast}$ & $\hat{b}_{\mu}$ & $\approx \Xi_b \hat{b}_{\ast\mu}$ \\
$\hat{d}_{\ast}$ & $d^{(d)}_{\mu\alpha_1\dots \alpha_{d-3}}u^{\mu}u^{\alpha_1}\dots u^{\alpha_{d-3}}$ & $m_{\psi}^{d-3}/(u^2)^{(d-3)/2}$ & $u^{\mu}\frac{\partial \hat{d}_{\ast}}{\partial u^{\mu}}$ & $=(d-2)\hat{d}_{\ast}$ & $\hat{d}_{\mu}$ & $\approx \Xi_d \hat{d}_{\ast\mu}$ \\
$\tilde{\hat{h}}_{\ast\mu}$ & $\tilde{H}^{(d)}_{\mu\nu\alpha_1\dots \alpha_{d-3}}u^{\nu}u^{\alpha_1}\dots u^{\alpha_{d-3}}$ & $m_{\psi}^{d-3}/(u^2)^{(d-3)/2}$ & $u^{\mu}\frac{\partial \tilde{\hat{h}}_{\ast\nu}}{\partial u^{\mu}}$ & $=(d-2)\tilde{\hat{h}}_{\ast\nu}$ & $\tilde{\hat{H}}_{\mu\nu}$ & $\approx \Xi_H\tilde{\hat{H}}_{\ast\mu\nu}$ \\
$\tilde{\hat{g}}_{\ast\mu}$ & $\tilde{g}^{(d)}_{\mu\nu\alpha_1\dots \alpha_{d-3}}u^{\nu}u^{\alpha_1}\dots u^{\alpha_{d-3}}$ & $m_{\psi}^{d-3}/(u^2)^{(d-3)/2}$ & $u^{\mu}\frac{\partial \tilde{\hat{g}}_{\ast\nu}}{\partial u^{\mu}}$ & $=(d-2)\tilde{\hat{g}}_{\ast\nu}$ & $\tilde{\hat{g}}_{\mu\nu}$ & $\approx\Xi_g\tilde{\hat{g}}_{\ast\mu\nu}$ & \\
\botrule
\end{tabular}} \\
\subfloat[]{\label{tab:functions-spin-nondegenerate-lagrangians}\begin{tabular}{ccrlrlc}
\toprule
$\hat{\mathscr{X}}_{\ast}$ & Explicit expression & \multicolumn{2}{c}{$\hat{X}_{\ast}$} & \multicolumn{2}{c}{Radial derivative} & Generalized bipartite matrix $s^{\hat{x}}_{\mu\nu}$ \\
\colrule
$\hat{\mathscr{B}}_{\ast}$ & $\sqrt{\hat{B}_{\ast}^2-(\hat{B}_{\ast}^{\mu}\hat{B}_{\ast\mu})u^2}$ & $\hat{B}_{\ast}$ & $=\Xi_bb_{\ast}$ & \phantom{hh}$u^{\mu}\frac{\partial \hat{\mathscr{B}}_{\ast}}{\partial u^{\mu}}$ & $=\hat{\mathscr{B}}_{\ast}$ & $\hat{B}_{\ast\mu}\hat{B}_{\ast\nu}-(\hat{B}_{\ast}^{\varrho}\hat{B}_{\ast\varrho})\eta_{\mu\nu}$ \\
$\hat{\mathscr{D}}_{\ast}$ & $\sqrt{\hat{D}_{\ast}^2-(\hat{D}_{\ast}^{\mu}\hat{D}_{\ast\mu})u^2}$ & $\hat{D}_{\ast}$ & $=\Xi_dd_{\ast}$ & $u^{\mu}\frac{\partial \hat{\mathscr{D}}_{\ast}}{\partial u^{\mu}}$ & $=\hat{\mathscr{D}}_{\ast}$ & $\hat{D}_{\ast\mu}\hat{D}_{\ast\nu}-(\hat{D}_{\ast}^{\varrho}\hat{D}_{\ast\varrho})\eta_{\mu\nu}$ \\
$\hat{\mathscr{H}}_{\ast}$ & $\sqrt{-\tilde{\hat{H}}_{\ast}^{\mu}\tilde{\hat{H}}_{\ast\mu}}$ & $\tilde{\hat{H}}_{\ast\mu}$ & $=\Xi_H\tilde{\hat{h}}_{\ast\mu}$ & $u^{\mu}\frac{\partial \hat{\mathscr{H}}_{\ast}}{\partial u^{\mu}}$ & $=\hat{\mathscr{H}}_{\ast}$ & $-\tilde{\hat{H}}^{\varrho}_{\ast}\tilde{\hat{H}}_{\ast\varrho}\eta_{\mu\nu}/u^2$ \\
$\hat{\mathscr{G}}_{\ast}$ & $\sqrt{-\tilde{\hat{G}}_{\ast}^{\mu}\tilde{\hat{G}}_{\ast\mu}}$ & $\tilde{\hat{G}}_{\ast\mu}$ & $=\Xi_g\tilde{\hat{g}}_{\ast\mu}$ & $u^{\mu}\frac{\partial \hat{\mathscr{G}}_{\ast}}{\partial u^{\mu}}$ & $=\hat{\mathscr{G}}_{\ast}$ & $-\tilde{\hat{G}}^{\varrho}_{\ast}\tilde{\hat{G}}_{\ast\varrho}\eta_{\mu\nu}/u^2$ \\
\botrule
\end{tabular}}
\caption{Ingredients necessary to formulate the classical Lagrangian of \eqref{eq:lagrangian-spin-nondegenerate} for a specific spin-nondegenerate operator. The first column of \protect\subref{tab:parameters-spin-nondegenerate-lagrangians} states the base observer scalars and vectors with their explicit construction given in the second column. The third column lists the consistency factors and the fourth gives Euler's theorem for the quantities defined before. In the fifth column, the reader can find the leading-order correspondences between the Lorentz-violating operators transformed to momentum space and the related parameters with all four-momenta replaced by four-velocities. The first column of \protect\subref{tab:functions-spin-nondegenerate-lagrangians} specifies the functions $\hat{\mathscr{X}}_{\ast}$ that make up the Lorentz-violating contribution of the Lagrangians. In the second column, the observer scalars and vectors necessary to construct these functions can be found. The third column points out Euler's theorem for each $\hat{\mathscr{X}}_{\ast}$ and in the fourth column the explicit matrices of the generalized bipartite Lagrangian of \eqref{eq:lagrangian-bipartite} are listed.}
\label{tab:parameters-spin-nondegenerate-lagrangians}
\end{table}

A similar procedure is applied to the $\hat{H}$ and $\hat{g}$ operators. Let $X\equiv H^{(3)}_{\mu\nu}H^{(3)\mu\nu}/4$ and $Y\equiv H^{(3)}_{\mu\nu}\tilde{H}^{(3)\mu\nu}/4$ with $\tilde{H}^{(3)\mu\nu}$ corresponding to the dual of $H^{(3)\mu\nu}$. The exact classical Lagrangian of the minimal $H$ coefficients for the configuration characterized by $Y=0$ is given by Eq.~(15) in \cite{Kostelecky:2010hs} and it reads
\begin{align}
L^{H^{(3)}}|_{Y=0}&=-m_{\psi}\sqrt{u^2}\mp\sqrt{u^{\nu}(H^{(3)})_{\nu}^{\phantom{\nu}\mu}H^{(3)}_{\mu\varrho}u^{\varrho}+2Xu^2} \notag \\
&=-m_{\psi}\sqrt{u^2}\mp\sqrt{-(\tilde{H}^{(3)})^{\mu}_{\phantom{\mu}\nu}u^{\nu}\tilde{H}^{(3)}_{\mu\varrho}u^{\varrho}}\,,
\end{align}
where we used
\begin{equation}
(H^{(3)})^{\mu}_{\phantom{\mu}\nu}u^{\nu}H^{(3)}_{\mu\varrho}u^{\varrho}=(\tilde{H}^{(3)})^{\mu}_{\phantom{\mu}\nu}u^{\nu}\tilde{H}^{(3)}_{\mu\varrho}u^{\varrho}+2Xu^2\,.
\end{equation}
Note that this Lagrangian is exact in Lorentz violation. Based on the same fundamental principles as before, we can propose a suitable \textit{Ansatz} to cover the $H$ coefficients for arbitrary mass dimension at first order in Lorentz violation:
\begin{subequations}
\begin{align}
L^{\hat{H}^{(d)}}&=-m_{\psi}\sqrt{u^2}\mp\hat{\mathscr{H}}_{\ast}+\dots\,,\quad \hat{\mathscr{H}}_{\ast}=\sqrt{-\tilde{\hat{H}}^{\mu}_{\ast}\tilde{\hat{H}}_{\ast\mu}}\,, \displaybreak[0]\\[2ex]
\tilde{\hat{H}}_{\ast\mu}&\equiv \Xi_h\tilde{\hat{h}}_{\ast\mu}\,,\quad \Xi_H=\Xi_H^{(d)}=\frac{m_{\psi}^{d-3}}{(u^2)^{(d-3)/2}}\,, \displaybreak[0]\\[2ex]
\tilde{\hat{h}}_{\ast\mu}&=\tilde{\hat{h}}^{(d)}_{\ast\mu}\equiv \tilde{H}^{(d)}_{\mu\nu\alpha_1\dots\alpha_{d-3}}u^{\nu}u^{\alpha_1}\dots u^{\alpha_{d-3}}\,.
\end{align}
\end{subequations}
The latter reproduces the minimal result. An analytical check of Eqs.~(\ref{eq:set-equations-lagrangians}) is presented in \appref{sec:analytical-proof-H}, i.e., the Lagrangian proposed is the correct first-order expression. It is valid for all configurations of $H$ coefficients, as a restriction  $Y=0$ generalized to $H^{(d)}_{\mu\nu}$ is not used in the proof. When taking into account that both the $H$ and the $g$ coefficients contribute to the two-tensor operator $\hat{\mathcal{T}}^{\mu\nu}$ in the Lagrange density of \eqref{eq:lagrange-density-fermion-sector}, the leading-order dispersion equation for $\hat{g}^{\mu\nu\varrho}$ follows from that of $\hat{H}^{\mu\nu}$ in replacing $\hat{H}^{\mu\nu}$ by $-\hat{g}^{\mu\nu\varrho}\overline{p}_{\varrho}$. The proof of the validity of the classical Lagrangian can then be adapted by considering $\hat{H}_{\ast\mu\nu}\leftrightarrow -\hat{g}_{\ast\mu\nu\varrho}u^{\varrho}=-\hat{g}_{\ast\mu\nu}$ at first order in the controlling coefficients. Note that the mass dimension is even and $\geq 4$ for $\hat{g}$.

Based on these findings, the general Lagrangian for a spin-nondegenerate operator at leading order is expressed as follows:
\begin{equation}
\label{eq:lagrangian-spin-nondegenerate}
L^{\hat{x}^{(d)}}=-m_{\psi}\sqrt{u^2}\mp \hat{\mathscr{X}}_{\ast}\,,
\end{equation}
with the quantities listed in \tabref{tab:parameters-spin-nondegenerate-lagrangians}.
Several remarks are in order. First, the standard Lagrangian is reproduced for vanishing controlling coefficients, as expected. Second, each Lorentz-violating coefficient is multiplied by a parameter $\Xi_x$ that only depends on the particle mass, the four-velocity squared, and the mass dimension. This parameter has the same form for each type of spin-nondegenerate operator. However, note that the mass dimensions of the coefficients can differ from each other. Third, Lorentz violation is encoded in a square root of quadratic combinations of coefficients, i.e., the correction is of first order in Lorentz violation. Due to the square root dependence, the Lagrangian is not differentiable in the limit of zero Lorentz violation, though. Fourth, there are two Lagrangians for particles that mirror the two distinct modified dispersion relations present for spin-nondegenerate operators. Fifth, taking a closer look at the Lagrangians reveals the correspondence $\tilde{\hat{H}}_{\ast\mu}\leftrightarrow m_{\psi}\hat{d}_{\ast\mu}$ for a $\hat{d}_{\mu\nu}$ that is antisymmetric in the first two indices (cf.~\cite{Russell:2015gwa} for the analog in the minimal SME). Sixth, at first order in Lorentz violation, the Lagrangians for $\hat{b}$, $\hat{d}$, $\hat{H}$, and $\hat{g}$ are all of the following form:
\begin{equation}
\label{eq:lagrangian-bipartite}
L^{\hat{x}^{(d)}}=-m_{\psi}\sqrt{u^2}\mp\sqrt{u^{\mu}s^{\hat{x}}_{\mu\nu}u^{\nu}}\,,
\end{equation}
where $s^{\hat{x}}_{\mu\nu}=s^{\hat{x}}_{\mu\nu}(u)$ are $4\times 4$ matrices that are listed in \tabref{tab:functions-spin-nondegenerate-lagrangians} explicitly.
In principle, \eqref{eq:lagrangian-bipartite} can be interpreted as a generalization of what is known as the bipartite structure within the minimal SME \cite{Kostelecky:2012ac}. However, in contrast to its original definition, the matrix $s^{\hat{x}}_{\mu\nu}$ now depends on the four-velocity explicitly. Note the formal similarities of these matrices for all types of spin-nondegenerate operators when taking into account that $\tilde{\hat{H}}_{\ast}^{\mu}u_{\mu}=\tilde{\hat{g}}_{\ast}^{\mu}u_{\mu}=0$ because of the antisymmetry of $\tilde{\hat{H}}$ and $\tilde{\hat{g}}$. Seventh, the Lagrangians for the first two types and the latter two types of operators can be combined resulting in Lagrangians expressed in terms of an observer pseudovector and a two-tensor that are defined in analogy to the pseudovector operator $\hat{\mathcal{A}}^{\mu}$ and the dual tensor operator $\tilde{\hat{\mathcal{T}}}^{\mu\nu}$:
\begin{subequations}
\begin{align}
\label{eq:lagrangian-curly-ahat}
L^{\hat{\mathcal{A}}^{(d)}}|_{\substack{d\geq 3 \\
\text{even}}}&=-m_{\psi}\sqrt{u^2}\mp\sqrt{\hat{\mathcal{A}}_{\ast}^2-(\hat{\mathcal{A}}_{\ast}^{\mu}\hat{\mathcal{A}}_{\ast\mu})u^2}\,, \displaybreak[0]\\[2ex]
\label{eq:lagrangian-curly-that}
L^{\tilde{\hat{\mathcal{T}}}^{(d)}}|_{\substack{d\geq 3 \\
\text{odd}}}&=-m_{\psi}\sqrt{u^2}\mp\sqrt{-\tilde{\hat{\mathcal{T}}}^{\mu}_{\ast}\tilde{\hat{\mathcal{T}}}_{\ast\mu}}\,, \displaybreak[0]\\[2ex]
\hat{\mathcal{A}}_{\ast\mu}^{(d)}&\equiv\Xi_b^{(d)}\left(\frac{m_{\psi}}{\sqrt{u^2}}\hat{d}_{\ast\mu}^{(d+1)}-\hat{b}_{\ast\mu}^{(d)}\right)\,,\quad \tilde{\hat{\mathcal{T}}}_{\ast\mu}^{(d)}\equiv\Xi_H^{(d)}\left(\frac{m_{\psi}}{\sqrt{u^2}}\tilde{\hat{g}}_{\ast\mu}^{(d+1)}-\tilde{\hat{H}}_{\ast\mu}^{(d)}\right)\,.
\end{align}
\end{subequations}
Here, the index indicating the mass dimension of $\hat{\mathcal{A}}_{\ast\mu}$, $\tilde{\hat{\mathcal{T}}}_{\ast\mu}$ is again omitted within the Lagrangians to simplify the notation. The Lagrangians for $\hat{b}$, $\hat{d}$ follow from \eqref{eq:lagrangian-curly-ahat} and those for $\hat{H}$, $\hat{g}$ follow from \eqref{eq:lagrangian-curly-that} for appropriate choices of the coefficients, as expected. These new Lagrangians are of bipartite form, as well. Last, but not least, an alternative possibility is to express the Lagrangians in terms of effective coefficients based on Eqs.~(27) of \cite{Kostelecky:2013rta}:
\begin{subequations}
\begin{align}
\label{eq:lagrangian-geff}
L^{\tilde{\hat{g}}^{(d)}_{\mathrm{eff}}}|_{\substack{d\geq 4 \\
\text{even}}}&=-m_{\psi}\sqrt{u^2}\mp \sqrt{-\tilde{\hat{G}}^{\mu}_{\ast}\tilde{\hat{G}}_{\ast\mu}+\frac{2}{m_{\psi}}\left[\tilde{\hat{G}}_{\ast}\hat{B}_{\ast}-(\tilde{\hat{G}}^{\mu}_{\ast}\hat{B}_{\ast\mu})u^2\right]+\frac{u^2}{m_{\psi}^2}\left[\hat{B}_{\ast}^2-(\hat{B}^{\mu}_{\ast}\hat{B}_{\ast\mu})u^2\right]} \notag \\
&=-m_{\psi}\sqrt{u^2}\mp\sqrt{-\tilde{\hat{G}}_{\mathrm{eff},\ast}^{\mu}\tilde{\hat{G}}_{\mathrm{eff},\ast\mu}}\,, \displaybreak[0]\\[2ex]
\label{eq:lagrangian-Heff}
L^{\tilde{\hat{H}}^{(d)}_{\mathrm{eff}}}|_{\substack{d\geq 3 \\
\text{odd}}}&=-m_{\psi}\sqrt{u^2}\mp\sqrt{-\tilde{\hat{H}}^{\mu}_{\ast}\tilde{\hat{H}}_{\ast\mu}+\frac{2}{m_{\psi}}\left[\tilde{\hat{H}}_{\ast}\hat{D}_{\ast}-(\tilde{\hat{H}}^{\mu}_{\ast}\hat{D}_{\ast\mu})u^2\right]+\frac{u^2}{m_{\psi}^2}\left[\hat{D}_{\ast}^2-(\hat{D}^{\mu}_{\ast}\hat{D}_{\ast\mu})u^2\right]} \notag \\
&=-m_{\psi}\sqrt{u^2}\mp\sqrt{-\tilde{\hat{H}}_{\mathrm{eff},\ast}^{\mu}\tilde{\hat{H}}_{\mathrm{eff},\ast\mu}}\,, \displaybreak[0]\\[2ex]
\tilde{\hat{G}}_{\ast\mu,\mathrm{eff}}^{(d)}&\equiv\Xi^{(d)}_g\tilde{\hat{g}}_{\ast\mu,\mathrm{eff}}^{(d)}\,,\quad\tilde{\hat{g}}_{\ast\mu,\mathrm{eff}}^{(d)}\equiv \tilde{g}^{(d)}_{\mathrm{eff},\mu\nu\alpha_1\dots\alpha_{d-3}}u^{\nu}u^{\alpha_1}\dots u^{\alpha_{d-3}}\,, \displaybreak[0]\\[2ex]
\tilde{\hat{H}}_{\ast\mu,\mathrm{eff}}^{(d)}&\equiv\Xi^{(d)}_H\tilde{\hat{h}}_{\ast\mu,\mathrm{eff}}^{(d)}\,,\quad\tilde{\hat{h}}_{\ast\mu,\mathrm{eff}}^{(d)}\equiv \tilde{H}^{(d)}_{\mathrm{eff},\mu\nu\alpha_1\dots\alpha_{d-3}}u^{\nu}u^{\alpha_1}\dots u^{\alpha_{d-3}}\,.
\end{align}
\end{subequations}
In Eqs.~(\ref{eq:lagrangian-geff}), (\ref{eq:lagrangian-Heff}) the mass dimension of the effective coefficients has again been omitted for brevity. The Lagrangians for the $\hat{b}$, $\hat{g}$ operators are contained in \eqref{eq:lagrangian-geff} as special cases where those for $\hat{d}$, $\hat{H}$ follow from \eqref{eq:lagrangian-Heff} by setting suitable coefficients to zero. Note that the pairs of coefficients even mix in the Lagrangians of this form.

Finally, the proofs for the operators $\hat{b}$, $\hat{H}$ shown in Appx.~\ref{sec:analytical-proof-b}, \ref{sec:analytical-proof-H} can be taken over literally to the situation when operators of different mass dimensions are summed over. The argument is the same as that presented for the spin-degenerate operators at the end of \secref{sec:results-spin-degenerate-operators}. The only thing to do is to replace $\hat{X}_{\ast}^{(d)}$ by a suitable sum, i.e., $\hat{X}_{\ast}^{(d)}\mapsto \sum_d \hat{X}_{\ast}^{(d)}$ where the summation runs over all $d$ permitted. The latter sum is then still positively homogeneous of degree 1 in the velocity, as the individual contributions are. Note that $\hat{\mathscr{X}}_{\ast}$ depends on bilinear combinations of $\hat{X}_{\ast}$, i.e., summations over the mass dimension are carried out under the square root and not in front of it. By doing so, coefficients of different mass dimensions may mix.

\section{General first-order Lagrangian of SME fermion sector}
\label{sec:general-first-order-lagrangian}

Comparing the previously obtained Lagrangians and the corresponding first-order dispersion relations reveals plenty of similarities. Therefore, we found that there exists a direct map from the dispersion relation $E^{(\pm)}$ to the associated classical Lagrangians $L^{(\pm)}$ at first order in Lorentz violation. Consider the first-order dispersion relation of the nonminimal SME that is known to be of the form
\begin{subequations}
\label{eq:first-order-dispersion-relation}
\begin{align}
E^{(\pm)}&=E_0-\frac{1}{E_0}\left(p\cdot\hat{\mathcal{V}}_{\mathrm{eff}}\mp\Upsilon\right)\,, \displaybreak[0]\\[2ex]
\Upsilon&=\sqrt{p^{\mu}(\tilde{\hat{\mathcal{T}}}_{\mathrm{eff}})_{\mu\varrho}(\tilde{\hat{\mathcal{T}}}_{\mathrm{eff}})^{\varrho}_{\phantom{\varrho}\nu}p^{\nu}}\,,
\end{align}
with the effective operators transformed to momentum space
\begin{equation}
\hat{\mathcal{V}}^{\mu}_{\mathrm{eff}}\equiv\hat{\mathcal{V}}^{\mu}+\frac{1}{m_{\psi}}p^{\mu}\hat{\mathcal{S}}\,,\quad \tilde{\hat{\mathcal{T}}}_{\mathrm{eff}}^{\mu\nu}\equiv\tilde{\hat{\mathcal{T}}}^{\mu\nu}+\frac{1}{m_{\psi}}p^{[\mu}\hat{\mathcal{A}}^{\nu]}\,,
\end{equation}
\end{subequations}
cf.~Eq.~(43) in \cite{Kostelecky:2013rta}. The map leading directly from the dispersion relation to the classical Lagrangian involves the following steps. First, perform the replacement $E_0\mapsto -m_{\psi}\sqrt{u^2}$. Second, carry out $p_{\mu}\mapsto m_{\psi}u_{\mu}/\sqrt{u^2}$ in the Lorentz-violating term. Third, multiply the Lorentz-violating contribution by $u^2$ to ensure positive homogeneity of first degree in the velocity. An analog map was found in~\cite{Schreck:2016jqn} within the minimal SME where its validity was demonstrated to second order in the velocity and momentum only. The procedure previously described is a generalization that is valid in the nonminimal SME and at all orders in the velocity and momentum. Applying this map to the dispersion relation of \eqref{eq:first-order-dispersion-relation}, produces the first-order classical ``master'' Lagrangian including all operators of the nonminimal SME:
\begin{subequations}
\begin{align}
L^{(\pm)}_{\mathrm{master}}&=-m_{\psi}\sqrt{u^2}+\hat{\mathcal{V}}_{\ast,\mathrm{eff}}\mp\Upsilon_{\ast}\,, \displaybreak[0]\\[2ex]
\Upsilon_{\ast}&=\sqrt{-(\tilde{\hat{\mathcal{T}}}_{\ast,\mathrm{eff}})^{\mu}(\tilde{\hat{\mathcal{T}}}_{\ast,\mathrm{eff}})_{\mu}}\,,
\end{align}
with a slew of observer scalars and (pseudo)vectors that correspond to the effective operators considered:
\begin{align}
\hat{\mathcal{V}}_{\ast,\mathrm{eff}}&\equiv\hat{\mathcal{V}}_{\ast}+\sqrt{u^2}\hat{\mathcal{S}}_{\ast}\,, \\[2ex]
\tilde{\hat{\mathcal{T}}}_{\ast,\mathrm{eff}}^{\mu}&\equiv\tilde{\hat{\mathcal{T}}}_{\ast}^{\mu}+\frac{u^{\mu}}{\sqrt{u^2}}\hat{\mathcal{A}}_{\ast}-\sqrt{u^2}\hat{\mathcal{A}}_{\ast}^{\mu}\,, \displaybreak[0]\\[2ex]
\hat{\mathcal{S}}_{\ast}&\equiv\sum_{d\geq 4,\text{even}} \Xi_{\mathcal{V}}^{(d)}\left(\hat{e}_{\ast}^{(d)}-\frac{m_{\psi}}{\sqrt{u^2}}\hat{m}_{\ast}^{(d+1)}\right)\,, \displaybreak[0]\\[2ex]
\hat{\mathcal{V}}_{\ast\mu}&\equiv\sum_{d\geq 3,\text{odd}} \Xi_{\mathcal{V}}^{(d)}\left(\frac{m_{\psi}}{\sqrt{u^2}}\hat{c}^{(d+1)}_{\ast\mu}-\hat{a}_{\ast\mu}^{(d)}\right)\,,\quad\hat{\mathcal{V}}_{\ast}\equiv\hat{\mathcal{V}}_{\ast\mu}u^{\mu}\,,\quad \Xi_{\mathcal{V}}^{(d)}=\frac{m_{\psi}^{d-3}}{(u^2)^{(d-3)/2}}\,, \displaybreak[0]\\[2ex]
\hat{\mathcal{A}}_{\ast\mu}&\equiv\sum_{d\geq 3,\text{odd}} \Xi_{\mathcal{A}}^{(d)}\left(\frac{m_{\psi}}{\sqrt{u^2}}\hat{d}_{\ast\mu}^{(d+1)}-\hat{b}_{\ast\mu}^{(d)}\right)\,,\quad\hat{\mathcal{A}}_{\ast}\equiv\hat{\mathcal{A}}_{\ast\mu}u^{\mu}\,,\quad \Xi_{\mathcal{A}}^{(d)}=\frac{m_{\psi}^{d-3}}{(u^2)^{(d-3)/2}}\,, \displaybreak[0]\\[2ex]
\tilde{\hat{\mathcal{T}}}_{\ast\mu}&\equiv\sum_{d\geq 3,\text{odd}} \Xi_{\mathcal{A}}^{(d)}\left(\frac{m_{\psi}}{\sqrt{u^2}}\tilde{\hat{g}}_{\ast\mu}^{(d+1)}-\tilde{\hat{H}}_{\ast\mu}^{(d)}\right)\,.
\end{align}
\end{subequations}
Several remarks are in order. First, the Lagrangians associated to the spin-degenerate operators are completely governed by $\hat{\mathcal{V}}_{\ast,\mathrm{eff}}$, whereas the spin-nondegenerate results are described by $\tilde{\hat{\mathcal{T}}}_{\ast,\mathrm{eff}}$. Second, all Lagrangians found previously are contained in the latter general result, which can be checked by setting subsets of the coefficients to zero. Third, the two signs before the spin-nondegenerate contribution are switched when comparing the dispersion relations to the Lagrangians. Fourth, the only Lagrangian that is not directly contained in this general result is that for the $f$ coefficients. Since the Lagrangian, as it stands, it valid at first order in Lorentz violation only, we do not intend to add the Lagrangian for $\hat{f}$, as the latter coefficients do not deliver a linear contribution. However, recall that $f_{\ast}^{(d)}$ squared can just be mapped onto $c^{(d)}_{\ast}$, cf.~Eqs.~(\ref{eq:correspondence-c-f}).

The proof that the latter Lagrangian fulfills the defining equations (\ref{eq:set-equations-lagrangians}) can be put together from the proofs previously carried out. The Lagrangian for $\tilde{\hat{\mathcal{T}}}_{\ast,\mathrm{eff}}=0$ is that of the spin-degenerate operators making the corresponding proof of \secref{sec:results-spin-degenerate-operators} applicable. For $\hat{\mathcal{V}}_{\ast,\mathrm{eff}}=0$ we can take over the proof for $\hat{H}$ of \appref{sec:analytical-proof-H}, as the Lagrangian for the $H$ coefficients is exactly of this form. Last but not least, each contribution is of first order in Lorentz violation, which is why both proofs can be combined. The spin-degenerate and spin-nondegenerate operators do not mix with each other at this level of approximation, after all.

\section{Conclusions}
\label{sec:conclusions}

In the current paper, we derived the leading-order classical Lagrangian covering all operators of the nonminimal SME fermion sector. The result for a particular operator can be obtained from this master Lagrangian by setting all other coefficients to zero. The Lagrangian for each operator was found to be a natural generalization of the already known minimal results where the minimal coefficients are replaced by the infinite sum over all nonminimal coefficients appropriately contracted with four-velocities. Furthermore, the Lorentz-violating contributions are multiplied by a factor ensuring both consistency of the mass dimension and positive homogeneity of first degree in the velocity. The first-order Lagrangian shares a lot of similarities with the corresponding dispersion relations. The modified terms for the spin-degenerate operators are directly proportional to sums of the coefficients suitably contracted with four-velocities. The Lorentz-violating terms linked to the spin-nondegenerate operators involve a square root of a bilinear combination of controlling coefficients and four-velocities. Terms of these shapes can be considered as generalizations of structures that are known as bipartite in the literature.

With the master Lagrangian at hand, the description of Lorentz violation for classical systems should now be feasible for any kind of Lorentz-violating operator. The only caveat is that the Lagrangian obtained is the leading-order result only. However, as Lorentz violation is perturbative, the result should at least be sufficient for phenomenological studies. Terms of higher order in the perturbative expansion are supposedly much more involved, as suitable observer scalars can now be formed from component coefficients of different type and mass dimension. In future works, we intend to investigate whether the Lagrangians found can be promoted to Finsler structures. If this turns out to be possible, the properties of these Finsler structures will be an interesting topic to investigate both for physicists and mathematicians.

\section{Acknowledgments}

It is a pleasure to thank V.A.~Kosteleck\'{y} for suggestions on an early draft of the paper. The authors are indebted to FAPEMA for financial support via the grant FAPEMA/Pos-grad/03978/15.

\newpage
\begin{appendix}
\numberwithin{equation}{section}

\section{Analytical proof for $b$ coefficients}
\label{sec:analytical-proof-b}
\setcounter{equation}{0}

In this part of the appendix, we would like to demonstrate that the first-order classical Lagrangian of \eqref{eq:classical-lagrangian-operator-b} satisfies the defining equations of the map from the field-theory description to the classical point-particle analog. The calculation will be carried out by analytical means where higher-order terms will be discarded. The Lorentz-violating contribution is positively homogeneous of first degree in the velocity, which will turn out to be very helpful. Before starting with the proof, recall the classical Lagrangian found for the $b$ coefficients:
\begin{subequations}
\begin{align}
L^{\hat{b}^{(d)}}&=-m_{\psi}\sqrt{u^2}\mp \hat{\mathscr{B}}_{\ast}\,, \\[2ex]
\label{eq:definition-curly-B}
\hat{\mathscr{B}}_{\ast}&=\sqrt{\hat{B}_{\ast}^2-(\hat{B}_{\ast}^{\mu}\hat{B}_{\ast\mu})u^2}\,.
\end{align}
\end{subequations}

\subsection{Dispersion equation}

The first and undoubtedly easier task is to check the validity of \eqref{eq:dispersion-relation} for the canonical momentum. The latter can be cast into the form
\begin{equation}
p_{\mu}=-\frac{\partial L^{\hat{b}^{(d)}}}{\partial u^{\mu}}=m_{\psi}\frac{u_{\mu}}{\sqrt{u^2}}\pm\frac{\partial\hat{\mathscr{B}}_{\ast}}{\partial u^{\mu}}\,.
\end{equation}
In the Lorentz-violating operator it suffices to replace each momentum by the leading-order term of the previous expression such that $\hat{b}_{\mu}\approx\Xi_b\hat{b}_{\ast\mu}$.
The general dispersion equation has the form
\begin{equation}
\mathcal{R}=(p^2-m_{\psi}^2)^2+4\left[\hat{b}^2m_{\psi}^2-(\hat{b}\cdot p)^2\right]+\dots\,.
\end{equation}
The leading-order Lorentz-violating terms in the dispersion equation are of second order, which is why these contributions have to cancel each other when the classical Lagrangian is supposed to be valid. Hence, all expressions must be computed at second order in Lorentz violation where higher-order terms are discarded. The square of the canonical momentum is given by
\begin{equation}
p^2=m_{\psi}^2\pm \frac{2m_{\psi}}{\sqrt{u^2}}u^{\mu}\frac{\partial\hat{\mathscr{B}}_{\ast}}{\partial u^{\mu}}+\left(\frac{\partial\hat{\mathscr{B}}_{\ast}}{\partial u}\right)^2=m_{\psi}^2\pm\frac{2m_{\psi}}{\sqrt{u^2}}\hat{\mathscr{B}}_{\ast}+\left(\frac{\partial\hat{\mathscr{B}}_{\ast}}{\partial u}\right)^2\,,
\end{equation}
where we used Euler's theorem for $\hat{\mathscr{B}}_{\ast}$ in the form
\begin{equation}
\label{eq:euler-theorem-bstar}
u^{\mu}\frac{\partial\hat{\mathscr{B}}_{\ast}}{\partial u^{\mu}}=\hat{\mathscr{B}}_{\ast}\,.
\end{equation}
The remaining terms in the dispersion equation at second order in Lorentz violation are written as follows:
\begin{subequations}
\begin{align}
\hat{b}\cdot p&=m_{\psi}\frac{\hat{B}_{\ast}}{\sqrt{u^2}}\pm\hat{B}_{\ast}^{\mu}\frac{\partial \hat{\mathscr{B}}_{\ast}}{\partial u^{\mu}}+\dots\,, \displaybreak[0]\\[2ex]
(\hat{b}\cdot p)^2&=m_{\psi}^2\frac{\hat{B}_{\ast}^2}{u^2}+\dots\,.
\end{align}
\end{subequations}
Inserting these results into the dispersion equation, all of the second-order terms in Lorentz violation compensate each other:
\begin{align}
\mathcal{R}&=\left(\pm\frac{2m_{\psi}}{\sqrt{u^2}}\hat{\mathscr{B}}_{\ast}\right)^2+4\left(\hat{b}^2m_{\psi}^2-m_{\psi}^2\frac{\hat{B}_{\ast}^2}{u^2}\right)+\dots \notag \\
&=\frac{4m_{\psi}^2}{u^2}\hat{\mathscr{B}}_{\ast}^2+4m_{\psi}^2\left(\hat{B}_{\ast}^{\mu}\hat{B}_{\ast\mu}-\frac{\hat{B}_{\ast}^2}{u^2}\right)=0\,,
\end{align}
according to \eqref{eq:definition-curly-B}. This outcome demonstrates that the canonical momentum based on the classical Lagrangian satisfies the dispersion equation at the order desired.

\subsection{Velocity correspondence}

In this paragraph we would like to demonstrate the validity of \eqref{eq:group-velocity-correspondence} for the classical Lagrangian under consideration. As the general formula for the dispersion relation may be complicated, we compute the first (implicit) derivative of the dispersion equation and replace all $\partial p_0/\partial p_i$ by $-u^i/u^0$. For the classical Lagrangian to be valid, all contributions at second order in Lorentz violation have to compensate each other. It is reasonable to split the implicit derivative into three parts as follows:
\begin{subequations}
\begin{align}
\frac{\partial\mathcal{R}}{\partial p_i}&=\sum_{i=1\dots 3} \left.\frac{\partial\mathcal{R}}{\partial p_i}\right|^{(i)}\,, \displaybreak[0]\\[2ex]
\left.\frac{\partial\mathcal{R}}{\partial p_i}\right|^{(1)}&=-4(p^2-m_{\psi}^2)\left(\frac{u^i}{u^0}p_0+p_i\right)\,, \displaybreak[0]\\[2ex]
\left.\frac{\partial\mathcal{R}}{\partial p_i}\right|^{(2)}&=8(\hat{b}\cdot p)\left(\frac{u^i}{u^0}\hat{b}_0+\hat{b}_i\right)\,, \displaybreak[0]\\[2ex]
\left.\frac{\partial\mathcal{R}}{\partial p_i}\right|^{(3)}&=8m_{\psi}^2\hat{b}^{\nu}\frac{\partial\hat{b}_{\nu}}{\partial p_i}-8(\hat{b}\cdot p)p^{\nu}\frac{\partial \hat{b}_{\nu}}{\partial p_i}\,.
\end{align}
\end{subequations}
The third contribution takes into account a possible momentum dependence of the controlling coefficients that arises for nonminimal frameworks. For the minimal $b$ coefficients, this term just vanishes. We start computing the first part:
\begin{equation}
\left.\frac{\partial\mathcal{R}}{\partial p_i}\right|^{(1)}=-\frac{8m_{\psi}}{\sqrt{u^2}}\hat{\mathscr{B}}_{\ast}\left(\frac{u^i}{u^0}\frac{\partial \hat{\mathscr{B}}_{\ast}}{\partial u^0}+\frac{\partial \hat{\mathscr{B}}_{\ast}}{\partial u^i}\right)+\dots\,.
\end{equation}
The second part of the implicit derivative can be obtained quickly, as well:
\begin{equation}
\left.\frac{\partial\mathcal{R}}{\partial p_i}\right|^{(2)}=\frac{8m_{\psi}}{\sqrt{u^2}}(\hat{b}\cdot u)\left(\frac{u^i}{u^0}\hat{b}_0+\hat{b}_i\right)+\dots=\frac{8m_{\psi}}{\sqrt{u^2}}\hat{B}_{\ast}\left(\frac{u^i}{u^0}\hat{B}_{\ast 0}+\hat{B}_{\ast i}\right)\,.
\end{equation}
Summing the two contributions obtained leads to
\begin{align}
\sum_{i=1\dots 2}\left.\frac{\partial\mathcal{R}}{\partial p_i}\right|^{(i)}&=\frac{8m_{\psi}}{\sqrt{u^2}}\left[
\hat{B}_{\ast}\left(\frac{u^i}{u^0}\hat{B}_{\ast 0}+\hat{B}_{\ast i}\right)-\hat{\mathscr{B}}_{\ast}\left(\frac{u^i}{u^0}\frac{\partial \hat{\mathscr{B}}_{\ast}}{\partial u^0}+\frac{\partial \hat{\mathscr{B}}_{\ast}}{\partial u^i}\right)\right]+\dots \notag \displaybreak[0]\\
&=\frac{8m_{\psi}}{\sqrt{u^2}}\left\{\hat{B}_{\ast}\left(\frac{u^i}{u^0}\hat{B}_{\ast 0}+\hat{B}_{\ast i}\right)\right. \notag \\
&\phantom{{}={}\frac{8m_{\psi}}{\sqrt{u^2}}\Bigg[}\left.-\left[\frac{u^i}{u^0}\left(\hat{B}_{\ast}\frac{\partial\hat{B}_{\ast}}{\partial u^0}-u^2\hat{B}_{\ast}^{\nu}\frac{\partial\hat{B}_{\ast \nu}}{\partial u^0}\right)+\hat{B}_{\ast}\frac{\partial\hat{B}_{\ast}}{\partial u^i}-u^2\hat{B}_{\ast}^{\nu}\frac{\partial\hat{B}_{\ast \nu}}{\partial u^i}\right]\right\}\,,
\end{align}
where we used
\begin{equation}
\frac{\partial \hat{\mathscr{B}}_{\ast}}{\partial u^{\mu}}=\frac{1}{\hat{\mathscr{B}}_{\ast}}\left(\hat{B}_{\ast}\frac{\partial\hat{B}_{\ast}}{\partial u^{\mu}}-\hat{B}^{\nu}_{\ast}\hat{B}_{\ast \nu}u_{\mu}-u^2\hat{B}_{\ast}^{\nu}\frac{\partial\hat{B}_{\ast \nu}}{\partial u^{\mu}}\right)\,.
\end{equation}
Last but not least, the third part delivers
\begin{equation}
\left.\frac{\partial\mathcal{R}}{\partial p_i}\right|^{(3)}=8m_{\psi}^2\left(\hat{B}_{\ast}^{\nu}-\frac{\hat{B}_{\ast}u^{\nu}}{u^2}\right)\frac{\partial\hat{b}_{\nu}}{\partial p_i}+\dots\,.
\end{equation}
At this point we express the derivative of the controlling coefficients with respect to the momentum as a derivative with respect to the velocity:
\begin{equation}
\frac{\partial\hat{b}_{\mu}}{\partial p_i}=\frac{\partial\hat{b}_{\mu}}{\partial u^{\nu}}\frac{\partial u^{\nu}}{\partial p_i}=-\frac{\sqrt{u^2}}{m_{\psi}}\left(\frac{u^i}{u^0}\frac{\partial\hat{B}_{\ast\mu}}{\partial u^0}+\frac{\partial\hat{B}_{\ast\mu}}{\partial u^i}\right)\,,
\end{equation}
where we employed Eqs.~(\ref{eq:derivatives-four-velocity}), (\ref{eq:derivative-function-f-u2}). Finally, all second-order terms in Lorentz violation compensate each other in the implicit derivative:
\begin{align}
\frac{\partial\mathcal{R}}{\partial p_i}&=\frac{8m_{\psi}}{\sqrt{u^2}}\left[\hat{B}_{\ast}\left(\frac{u^i}{u^0}\hat{B}_{\ast 0}+\hat{B}_{\ast i}\right)-\hat{B}_{\ast}\left(\frac{u^i}{u^0}\frac{\partial\hat{B}_{\ast}}{\partial u^0}+\frac{\partial\hat{B}_{\ast}}{\partial u^i}\right)+u^2\left(\frac{u^i}{u^0}\hat{B}_{\ast}^{\nu}\frac{\partial\hat{B}_{\ast\nu}}{\partial u^0}+\hat{B}^{\nu}_{\ast}\frac{\partial\hat{B}_{\ast\nu}}{\partial u^i}\right)\right. \notag \\
&\phantom{{}={}\frac{8m_{\psi}}{\sqrt{u^2}}\Big[}\left.-\left(\hat{B}_{\ast}^{\nu}u^2-\hat{B}_{\ast}u^{\nu}\right)\left(\frac{u^i}{u^0}\frac{\partial \hat{B}_{\ast\nu}}{\partial u^0}+\frac{\partial\hat{B}_{\ast\nu}}{\partial u^i}\right)\right]+\dots=0\,.
\end{align}
To arrive at this result, we additionally inserted
\begin{subequations}
\begin{align}
\frac{\partial\hat{B}_{\ast}}{\partial u^0}&=\frac{\partial(u^{\nu}\hat{B}_{\ast\nu})}{\partial u^0}=\hat{B}_{\ast}^0+u^{\nu}\frac{\partial\hat{B}_{\ast \nu}}{\partial u^0}\,, \\[2ex]
\frac{\partial\hat{B}_{\ast}}{\partial u^i}&=\frac{\partial(u^{\nu}\hat{B}_{\ast\nu})}{\partial u^i}=\hat{B}_{\ast i}+u^{\nu}\frac{\partial\hat{B}_{\ast \nu}}{\partial u^i}\,.
\end{align}
\end{subequations}

\section{Analytical proof for $H$ coefficients}
\label{sec:analytical-proof-H}
\setcounter{equation}{0}

Here we would like to carry out a proof analog to that for the $b$ coefficients. As a reminder, the classical Lagrangian found for the operator $\hat{H}$ is given by
\begin{subequations}
\begin{align}
L^{\hat{H}^{(d)}}&=-m_{\psi}\sqrt{u^2}\mp\hat{\mathscr{H}}_{\ast}\,, \\[2ex]
\label{eq:definition-curly-H}
\hat{\mathscr{H}}_{\ast}&=\sqrt{-\tilde{\hat{H}}_{\ast}^{\mu}\tilde{\hat{H}}_{\ast\mu}}\,.
\end{align}
\end{subequations}

\subsection{Dispersion equation}

First of all, it must be shown that the canonical momentum satisfies \eqref{eq:dispersion-relation}. Its form is completely analog to that for the $b$ coefficients:
\begin{equation}
p_{\mu}=-\frac{\partial L^{\hat{H}^{(d)}}}{\partial u^{\mu}}=m_{\psi}\frac{u_{\mu}}{\sqrt{u^2}}\pm\frac{\partial \hat{\mathscr{H}}_{\ast}}{\partial u^{\mu}}\,.
\end{equation}
Replacing the momentum components contracted with the $H$ coefficients by the standard term $m_{\psi}u^{\mu}/u^2$ produces $\hat{H}_{\mu\nu}\approx \Xi_H \hat{H}_{\ast\mu\nu}$. Neglecting all Lorentz-violating contributions beyond the second order, the dispersion equation can be expressed in the form
\begin{align}
\mathcal{R}&=p^4-2m_{\psi}^2p^2+m_{\psi}^4-m_{\psi}^2\hat{H}^{\mu\nu}\hat{H}_{\mu\nu}-2p_{\mu}(\hat{H}^{\mu\nu}-\mathrm{i}\tilde{\hat{H}}^{\mu\nu})(\hat{H}_{\nu\varrho}+\mathrm{i}\tilde{\hat{H}}_{\nu\varrho})p^{\varrho}+\dots \notag \\
&=(p^2-m_{\psi}^2)^2+(p^2-m_{\psi}^2)\hat{H}^{\mu\nu}\hat{H}_{\mu\nu}-4p^{\mu}\tilde{\hat{H}}_{\mu\varrho}\tilde{\hat{H}}^{\varrho}_{\phantom{\varrho}\nu}p^{\nu}+\dots \notag \\
&=(p^2-m_{\psi}^2)^2-4p^{\mu}\tilde{\hat{H}}_{\mu\varrho}\tilde{\hat{H}}^{\varrho}_{\phantom{\varrho}\nu}p^{\nu}+\dots\,,
\end{align}
where we used
\begin{equation}
p_{\mu}(\hat{H}^{\mu\nu}-\mathrm{i}\tilde{\hat{H}}^{\mu\nu})(\hat{H}_{\nu\varrho}+\mathrm{i}\tilde{\hat{H}}_{\nu\varrho})p^{\varrho}=2p^{\mu}\tilde{\hat{H}}_{\mu\varrho}\tilde{\hat{H}}^{\varrho}_{\phantom{\varrho}\nu}p^{\nu}-\frac{1}{2}p^2\hat{H}^{\mu\nu}\hat{H}_{\mu\nu}\,.
\end{equation}
The four-momentum squared reads
\begin{equation}
p^2=m_{\psi}^2\pm \frac{2m_{\psi}}{\sqrt{u^2}}u^{\mu}\frac{\partial \hat{\mathscr{H}}_{\ast}}{\partial u^{\mu}}+\left(\frac{\partial \hat{\mathscr{H}}_{\ast}}{\partial u}\right)^2=m_{\psi}^2\pm \frac{2m_{\psi}}{\sqrt{u^2}}\hat{\mathscr{H}}_{\ast}+\left(\frac{\partial \hat{\mathscr{H}}_{\ast}}{\partial u}\right)^2\,,
\end{equation}
where we used Euler's theorem applied to the characteristic quantity $\hat{\mathscr{H}}_{\ast}$:
\begin{equation}
u^{\mu}\frac{\partial \hat{\mathscr{H}}_{\ast}}{\partial u^{\mu}}=\hat{\mathscr{H}}_{\ast}\,.
\end{equation}
All of the ingredients are inserted into the dispersion equation demonstrating that the second-order terms in Lorentz violation cancel each other:
\begin{equation}
\mathcal{R}=\left(\pm \frac{2m_{\psi}}{\sqrt{u^2}}\hat{\mathscr{H}}_{\ast}\right)^2-\frac{4m_{\psi}^2}{u^2}\hat{\mathscr{H}}^2_{\ast}+\dots=0\,.
\end{equation}

\subsection{Velocity correspondence}

Now we demonstrate the validity of \eqref{eq:group-velocity-correspondence}. The first derivative of the dispersion equation is calculated implicitly with all $\partial p_0/\partial p_i$ replaced by $-u^i/u^0$. Due to the similarities of the dispersion equations for the $b$ and $H$ coefficients, we again split the derivative into three parts. The first two of these are obtained in total analogy to those for the $b$ coefficients:
\begin{align}
\left.\frac{\partial\mathcal{R}}{\partial p_i}\right|^{(1)}&\equiv \frac{\partial p^4}{\partial p_i}=-\frac{8m_{\psi}}{\sqrt{u^2}}\hat{\mathscr{H}}_{\ast}\left(\frac{u^i}{u^0}\frac{\partial \hat{\mathscr{H}}_{\ast}}{\partial u^0}+\frac{\partial \hat{\mathscr{H}}_{\ast}}{\partial u^i}\right)+\dots\,, \displaybreak[0]\\[2ex]
\left.\frac{\partial\mathcal{R}}{\partial p_i}\right|^{(2)}&\equiv 8(\tilde{\hat{H}}_{\varrho\mu}p^{\mu})\tilde{\hat{H}}^{\varrho\nu}\frac{\partial p_{\nu}}{\partial p_i}=\frac{8m_{\psi}}{\sqrt{u^2}}\tilde{\hat{H}}_{\varrho\mu}u^{\mu}\left(-\frac{u^i}{u^0}\tilde{\hat{H}}^{\varrho}_{\phantom{\varrho}0}-\tilde{\hat{H}}^{\varrho}_{\phantom{\varrho}i}\right) \notag \\
&=-\frac{8m_{\psi}}{\sqrt{u^2}}\tilde{\hat{H}}_{\ast\varrho}\left(\frac{u^i}{u^0}\tilde{\hat{H}}^{\varrho}_{\ast 0}+\tilde{\hat{H}}^{\varrho}_{\ast i}\right)+\dots\,.
\end{align}
Summing the latter contributions produces
\begin{align}
\sum_{i=1\dots 2}\left.\frac{\partial\mathcal{R}}{\partial p_i}\right|^{(i)}
&=-\frac{8m_{\psi}}{\sqrt{u^2}}\left[\tilde{\hat{H}}_{\ast\varrho}\left(\frac{u^i}{u^0}\tilde{\hat{H}}^{\varrho}_{\ast 0}+\tilde{\hat{H}}^{\varrho}_{\ast i}\right)
+\hat{\mathscr{H}}_{\ast}\left(\frac{u^i}{u^0}\frac{\partial \hat{\mathscr{H}}_{\ast}}{\partial u^0}+\frac{\partial \hat{\mathscr{H}}_{\ast}}{\partial u^i}\right)\right] \notag \\
&=\frac{8m_{\psi}}{\sqrt{u^2}}\tilde{\hat{H}}_{\ast\varrho}\left[\frac{u^i}{u^0}\left(\frac{\partial\tilde{\hat{H}}^{\varrho}_{\ast}}{\partial u^0}-\tilde{\hat{H}}^{\varrho}_{\ast 0}\right)+\frac{\partial \tilde{\hat{H}}_{\ast}^{\varrho}}{\partial u^i}-\tilde{\hat{H}}^{\varrho}_{\ast i}\right]+\dots\,,
\end{align}
where the derivative of the quantity $\hat{\mathscr{H}}_{\ast}$ was used:
\begin{equation}
\frac{\partial \hat{\mathscr{H}}_{\ast}}{\partial u^{\mu}}=-\frac{1}{\hat{\mathscr{H}}_{\ast}}\frac{\partial\tilde{\hat{H}}_{\ast}^{\nu}}{\partial u^{\mu}}\tilde{\hat{H}}_{\ast\nu}\,.
\end{equation}
The third part, which does not contribute to the minimal sector, contains derivatives of the Lorentz-violating operators with respect to the momentum:
\begin{align}
\left.\frac{\partial\mathcal{R}}{\partial p_i}\right|^{(3)}&=-4p^{\mu}\frac{\partial\tilde{\hat{H}}_{\mu}^{\phantom{\mu}\varrho}\tilde{\hat{H}}_{\varrho\nu}}{\partial p_i}p^{\nu}=-4\left(p^{\mu}\frac{\partial\tilde{\hat{H}}_{\mu}^{\phantom{\mu}\varrho}}{\partial p_i}\tilde{\hat{H}}_{\varrho\nu}p^{\nu}+p^{\mu}\tilde{\hat{H}}_{\mu\varrho}\frac{\partial\tilde{\hat{H}}^{\varrho}_{\phantom{\varrho}\nu}}{\partial p_i}p^{\nu}\right) \notag \displaybreak[0]\\
&=4\left\{\frac{\partial\tilde{\hat{H}}^{\varrho}}{\partial p^i}\tilde{\hat{H}}_{\varrho}-\left[\tilde{\hat{H}}^{\varrho}_{\phantom{\varrho}0}\left(-\frac{u^i}{u^0}\right)-\tilde{\hat{H}}^{\varrho}_{\phantom{\varrho} i}\right]\tilde{\hat{H}}_{\varrho}+\tilde{\hat{H}}_{\varrho}\frac{\partial\tilde{\hat{H}}^{\varrho}}{\partial p^i}-\tilde{\hat{H}}_{\varrho}\left[\tilde{\hat{H}}^{\varrho}_{\phantom{\varrho}0}\left(-\frac{u^i}{u^0}\right)-\tilde{\hat{H}}^{\varrho}_{\phantom{\varrho}i}\right]\right\} \notag \displaybreak[0]\\
&=8\tilde{\hat{H}}_{\varrho}\left[\frac{\partial\tilde{\hat{H}}^{\varrho}}{\partial p^i}+\frac{u^i}{u^0}\tilde{\hat{H}}^{\varrho}_{\phantom{\varrho}0}+\tilde{\hat{H}}^{\varrho}_{\phantom{\varrho} i}\right] \notag \displaybreak[0]\\
&=\frac{8m_{\psi}}{\sqrt{u^2}}\tilde{\hat{H}}_{\ast\varrho}\left[\frac{u^i}{u^0}\left(\tilde{\hat{H}}_{\ast 0}^{\varrho}-\frac{\partial\tilde{\hat{H}}_{\ast}^{\varrho}}{\partial u^0}\right)+\tilde{\hat{H}}_{\ast i}^{\varrho}-\frac{\partial\tilde{\hat{H}}_{\ast}^{\varrho}}{\partial u^i}\right]+\dots\,.
\end{align}
In the final step the derivative with respect to the momentum was again expressed as a derivative with respect to the four-velocity:
\begin{equation}
\frac{\partial\tilde{\hat{H}}_{\mu}}{\partial p_i}=\frac{\partial\tilde{\hat{H}}_{\mu}}{\partial u^{\sigma}}\frac{\partial u^{\sigma}}{\partial p_i}=-\left(\frac{u^i}{u^0}\frac{\partial\tilde{\hat{H}}_{\ast\mu}}{\partial u^0}+\frac{\partial\tilde{\hat{H}}_{\ast\mu}}{\partial u^i}\right)\,,
\end{equation}
where Eqs.~(\ref{eq:derivatives-four-velocity}), (\ref{eq:derivative-function-f-u2}) were employed in addition.
Now we see quickly that the sum of the first two parts equals the negative of the third, which makes all second-order terms in Lorentz violation vanish.

\end{appendix}

\newpage



\begin{thebibliography}{99}

\bibitem{Kostelecky:1988zi}
V.A.~Kosteleck\'{y} and S.~Samuel,
``Spontaneous breaking of Lorentz symmetry in string theory,''
Phys.\ Rev.\ D {\bf 39}, 683 (1989).

\bibitem{Kostelecky:1989jp}
V.A.~Kosteleck\'{y} and S.~Samuel,
``Phenomenological gravitational constraints on strings and higher-dimensional theories,''
Phys.\ Rev.\ Lett.\ {\bf 63}, 224 (1989).

\bibitem{Kostelecky:1989jw}
V.A.~Kosteleck\'{y} and S.~Samuel,
``Gravitational phenomenology in higher-dimensional theories and strings,''
Phys.\ Rev.\ D {\bf 40}, 1886 (1989).

\bibitem{Kostelecky:1991ak}
V.A.~Kosteleck\'{y} and R.~Potting,
``\textit{CPT} and strings,''
Nucl.\ Phys.\ B {\bf 359}, 545 (1991).

\bibitem{Kostelecky:1994rn}
V.A.~Kosteleck\'{y} and R.~Potting,
``\textit{CPT}, strings, and meson factories,''
Phys.\ Rev.\ D {\bf 51}, 3923 (1995),
hep-ph/9501341.

\bibitem{Gambini:1998it}
R.~Gambini and J.~Pullin,
``Nonstandard optics from quantum space-time,''
Phys.\ Rev.\ D {\bf 59}, 124021 (1999),
gr-qc/9809038.

\bibitem{Bojowald:2004bb}
M.~Bojowald, H.A.~Morales--T\'{e}cotl, and H.~Sahlmann,
``Loop quantum gravity phenomenology and the issue of Lorentz invariance,''
Phys.\ Rev.\ D {\bf 71}, 084012 (2005),
gr-qc/0411101.

\bibitem{AmelinoCamelia:1999pm}
G.~Amelino-Camelia and S.~Majid,
``Waves on noncommutative spacetime and gamma-ray bursts,''
Int.\ J.\ Mod.\ Phys.\ A {\bf 15}, 4301 (2000),
hep-th/9907110.

\bibitem{Carroll:2001ws}
S.M.~Carroll, J.A.~Harvey, V.A.~Kosteleck\'{y}, C.D.~Lane, and T.~Okamoto,
``Noncommutative field theory and Lorentz violation,''
Phys.\ Rev.\ Lett.\ {\bf 87}, 141601 (2001),
hep-th/0105082.

\bibitem{Klinkhamer:2003ec}
F.R.~Klinkhamer and C.~Rupp,
``Spacetime foam, \textit{CPT} anomaly, and photon propagation,''
Phys.\ Rev.\ D {\bf 70}, 045020 (2004),
hep-th/0312032.

\bibitem{Bernadotte:2006ya}
S.~Bernadotte and F.R.~Klinkhamer,
``Bounds on length scales of classical spacetime foam models,''
Phys.\ Rev.\ D {\bf 75}, 024028 (2007),
hep-ph/0610216.

\bibitem{Hossenfelder:2014hha}
S.~Hossenfelder,
``Theory and phenomenology of space-time defects,''
Adv.\ High Energy Phys.\ {\bf 2014}, 950672 (2014),
arXiv:1401.0276 [hep-ph].

\bibitem{Klinkhamer:1998fa}
F.R.~Klinkhamer,
``Z-string global gauge anomaly and Lorentz non-invariance,''
Nucl.\ Phys.\ B {\bf 535}, 233 (1998),
hep-th/9805095.

\bibitem{Klinkhamer:1999zh}
F.R.~Klinkhamer,
``A \textit{CPT} anomaly,''
Nucl.\ Phys.\ B {\bf 578}, 277 (2000),
hep-th/9912169.

\bibitem{Klinkhamer:2002mj}
F.R.~Klinkhamer and J.~Schimmel,
``{\em CPT} anomaly: a Rigorous result in four dimensions,''
Nucl.\ Phys.\ B {\bf 639}, 241 (2002),
hep-th/0205038.

\bibitem{Ghosh:2017iat}
K.J.B.~Ghosh and F.R.~Klinkhamer,
``Anomalous Lorentz and {\em CPT} violation from a local Chern-Simons-like term in the effective gauge-field action,''
Nucl.\ Phys.\ B {\bf 926}, 335 (2018),
arXiv:1706.07025 [hep-th].

\bibitem{Horava:2009uw}
P.~Ho\v{r}ava,
``Quantum gravity at a Lifshitz point,''
Phys.\ Rev.\ D {\bf 79}, 084008 (2009),
arXiv:0901.3775 [hep-th].

\bibitem{Colladay:1996iz}
D.~Colladay and V.A.~Kosteleck\'{y},
``{\em CPT} violation and the standard model,''
Phys.\ Rev.\ D {\bf 55}, 6760 (1997),
hep-ph/9703464.

\bibitem{Colladay:1998fq}
D.~Colladay and V.A.~Kosteleck\'{y},
``Lorentz-violating extension of the standard model,''
Phys.\ Rev.\ D {\bf 58}, 116002 (1998),
hep-ph/9809521.

\bibitem{Kostelecky:2003fs}
V.A.~Kosteleck\'{y},
``Gravity, Lorentz violation, and the standard model,''
Phys.\ Rev.\ D {\bf 69}, 105009 (2004)
hep-th/0312310].

\bibitem{Kostelecky:2009zp}
V.A.~Kosteleck\'{y} and M.~Mewes,
``Electrodynamics with Lorentz-violating operators of arbitrary dimension,''
Phys.\ Rev.\ D {\bf 80}, 015020 (2009),
arXiv:0905.0031 [hep-ph].

\bibitem{Kostelecky:2011gq}
V.A.~Kosteleck\'{y} and M.~Mewes,
``Neutrinos with Lorentz-violating operators of arbitrary dimension,''
Phys.\ Rev.\ D {\bf 85}, 096005 (2012),
arXiv:1112.6395 [hep-ph].

\bibitem{Kostelecky:2013rta}
V.A.~Kosteleck\'{y} and M.~Mewes,
``Fermions with Lorentz-violating operators of arbitrary dimension,''
Phys.\ Rev.\ D {\bf 88}, 096006 (2013),
arXiv:1308.4973 [hep-ph].

\bibitem{Greenberg:2002uu}
O.W.~Greenberg,
``\textit{CPT} violation implies violation of Lorentz invariance,''
Phys.\ Rev.\ Lett.\ {\bf 89}, 231602 (2002),
hep-ph/0201258.

\bibitem{Kostelecky:2008ts}
V.A.~Kosteleck\'{y} and N.~Russell,
``Data tables for Lorentz and \textit{CPT} violation,''
Rev.\ Mod.\ Phys.\  {\bf 83}, 11 (2011),
arXiv:0801.0287 [hep-ph].

\bibitem{Kostelecky:2010hs}
V.A.~Kosteleck\'{y} and N.~Russell,
``Classical kinematics for Lorentz violation,''
Phys.\ Lett.\ B {\bf 693}, 443 (2010)
arXiv:1008.5062 [hep-ph].

\bibitem{Colladay:2012rv}
D.~Colladay and P.~McDonald,
``Classical Lagrangians for momentum dependent Lorentz violation,''
Phys.\ Rev.\ D {\bf 85}, 044042 (2012),
arXiv:1201.3931 [hep-ph].

\bibitem{Russell:2015gwa}
N.~Russell,
``Finsler-like structures from Lorentz-breaking classical particles,''
Phys.\ Rev.\ D {\bf 91}, 045008 (2015),
arXiv:1501.02490 [hep-th].

\bibitem{Schreck:2014ama}
M.~Schreck,
``Classical kinematics for isotropic, minimal Lorentz-violating fermion operators,''
Phys.\ Rev.\ D {\bf 91}, 105001 (2015),
arXiv:1409.1539 [hep-th].

\bibitem{Schreck:2016jqn}
M.~Schreck,
``From classical Lagrangians to Hamilton operators in the Standard-Model Extension,''
Phys.\ Rev.\ D {\bf 94}, 025019 (2016),
arXiv:1602.01527 [hep-th].

\bibitem{Schreck:2015dsa}
M.~Schreck,
``Eikonal approximation, Finsler structures, and implications for Lorentz-violating photons in weak gravitational fields,''
Phys.\ Rev.\ D {\bf 92}, 125032 (2015),
arXiv:1508.00216 [hep-ph].

\bibitem{Schreck:2014hga}
M.~Schreck,
``Classical kinematics and Finsler structures for nonminimal Lorentz-violating fermions,''
Eur.\ Phys.\ J.\ C {\bf 75}, 187 (2015),
arXiv:1405.5518 [hep-th].

\bibitem{Schreck:2015seb}
M.~Schreck,
``Classical Lagrangians and Finsler structures for the nonminimal fermion sector of the Standard-Model Extension,''
Phys.\ Rev.\ D {\bf 93}, 105017 (2016),
arXiv:1512.04299 [hep-th].

\bibitem{Kostelecky:2011qz}
V.A.~Kosteleck\'{y},
``Riemann-Finsler geometry and Lorentz-violating kinematics,''
Phys.\ Lett.\ B {\bf 701}, 137 (2011),
arXiv:1104.5488 [hep-th].

%

\bibitem{Foster:2015yta}
J.~Foster and R.~Lehnert,
``Classical-physics applications for Finsler $b$ space,''
Phys.\ Lett.\ B {\bf 746}, 164 (2015),
arXiv:1504.07935 [physics.class-ph].

\bibitem{Colladay:2015wra}
D.~Colladay and P.~McDonald,
``Singular Lorentz-violating Lagrangians and associated Finsler structures,''
Phys.\ Rev.\ D {\bf 92}, 085031 (2015),
arXiv:1507.01018 [hep-ph].

\bibitem{Colladay:2017bon}
D.~Colladay,
``Extended Hamiltonian formalism and Lorentz-violating Lagrangians,''
Phys.\ Lett.\ B {\bf 772}, 694 (2017),
arXiv:1706.06637 [hep-th].

\bibitem{Kostelecky:2012ac}
V.A.~Kosteleck\'{y}, N.~Russell, and R.~Tso,
``Bipartite Riemann-Finsler geometry and Lorentz violation,''
Phys.\ Lett.\ B {\bf 716}, 470 (2012),
arXiv:1209.0750 [hep-th].

\bibitem{Altschul:2006ts}
B.~Altschul,
``Eliminating the {\em CPT}-odd $f$ coefficient from the Lorentz-violating standard model extension,''
J.\ Phys.\ A {\bf 39}, 13757 (2006),
hep-th/0602235.

\end{thebibliography}
\end{document}